\documentclass[journal]{IEEEtran}

\usepackage{cite}
\usepackage{amsmath}			
\usepackage{amssymb}			
\usepackage{verbatim}

\ifCLASSINFOpdf
   \usepackage[pdftex]{graphicx}
   \graphicspath{{figures/obliqueIncidence},{figures/pillboxMTS_LP},{figures/validation_examples},{figures}}
\else

\fi

\usepackage{amsmath}
\usepackage{amssymb}
\usepackage{bm}
\usepackage{relsize}
\usepackage{xcolor}

\usepackage{comment}
\usepackage{makecell}
\usepackage{float}
\usepackage{placeins}

\newcommand{\integSxDen}{{\mathop{\mathlarger{\int_{0}^{x} S(x') dx'}}}}
\newcommand{\integSxLDen}{{\mathop{\mathlarger{\int_{0}^{L} S(x) dx}}}}

\ifCLASSOPTIONcompsoc
  \usepackage[caption=false,font=normalsize,labelfont=sf,textfont=sf]{subfig}
\else
  \usepackage[caption=false,font=footnotesize]{subfig}
\fi

\begin{document}
%
\title{Plane-Wave Excitation of \\ Multi-Beam Modulated Metasurface Antennas}
%
%
%
\author{Jorge~Ruiz-Garc\'ia,
		Marco Faenzi,
		Adham Mahmoud,
		Mauro Ettorre,~\IEEEmembership{Fellow,~IEEE,}\\
		and~David~Gonz\'alez-Ovejero,~\IEEEmembership{Senior Member,~IEEE}
\thanks{The work of Marco Faenzi was supported in part by the F-NEW FRONTIERS (F-NF) 2024 Grant, through the Program Piano per lo Sviluppo della Ricerca 2024, issued by Italian Government and the University of Siena on May 17, 2024, under Project 95966. As for the rest of authors, they received the support of the European Union through the European Regional Development Fund (ERDF), and by the French Region of Brittany, Ministry of Higher Education and Research, Rennes M\'etropole and Conseil D\'epartemental 35, through the CPER Project STIC \& Ondes. It has also been supported by the Agence de l'Innovation de D\'{e}fense (AID), the \'Ecole des Docteurs de l'UBL (Université Bretagne Loire) and the Conseil R\'egional de Bretagne.}
\thanks{J. Ruiz-García and D. González-Ovejero are with the Institut d’Electronique et des Technologies du num\'eRique (IETR), UMR CNRS 6164, Universit\'e de Rennes 1, 35042 Rennes, France (e-mail: 
	jorge.ruiz-garcia@insa-rennes.fr).}%
\thanks{M. Faenzi is with the Department of Information Engineering and Mathematics, University of Siena, Italy.}
\thanks{A. Mahmoud and M. Ettorre were with the Institut d’Electronique et des Technologies du num\'eRique (IETR), UMR CNRS 6164, Universit\'e de Rennes 1, 35042 Rennes, France. A. Mahmoud is now with Isentroniq, 75006 Paris, France. M. Ettorre is now with the Electrical and Computer Engineering Department, Michigan State University, East Lansing, MI 48824 USA.}
}



\maketitle

\begin{abstract}
	This paper explores the design of multibeam metasurface (MTS) antennas excited by multi-directional plane-wave launchers. First, we solve the fundamental yet open problem of a plane surface wave (SW) that propagates obliquely to the modulation direction of a sinusoidally modulated MTS. Closed-form expressions are provided to accurately predict the beam pointing angles for any propagation direction of the illuminating plane SW and for any Floquet harmonic. Then, the proposed formulation is used to design a highly directive multibeam MTS antenna at K-band with linear polarization. The designed antenna combines an MTS and a pillbox quasi-optical beamformer arranged in a very compact space. A pillbox is a double-layer structure that embeds a reflector coupled to multiple primary feeds in the lower layer. The beams launched by the primary feeds are thus transformed into plane waves with different directions in the upper layer. Printing a modulated MTS in the upper layer results in a low-profile, multibeam antenna suitable for PCB fabrication. Experimental results validate the proposed formulation. The fabricated MTS antenna exhibits a maximum directivity of 30.5 dB and a 17.5\% fractional -3 dB directivity bandwidth over the 19.7-21.7 GHz band. By switching feeds and modifying the operating frequency, the scanning range is [3\textdegree,35\textdegree] in elevation and [-76\textdegree,+76\textdegree] in azimuth. Additionally, different antenna designs are included that demonstrate the validity and generality of the derived formulation. The proposed multibeam concept can be exploited in satellite communications and 5G/6G networks.
\end{abstract}

\begin{IEEEkeywords}
	Metasurface antennas, multibeam, plane-wave excitation, leaky waves, low profile, satellite communications, beamformers, PCB.
\end{IEEEkeywords}

%
\IEEEpeerreviewmaketitle

\section{Introduction}\label{sec:intro}
\bstctlcite{IEEEexample:BSTcontrol}

\IEEEPARstart{M}{odulated} metasurfaces (MTSs) are widely used for the design of high to very-high gain antennas \cite{Alu_Maci:2025,Zhao_Maci:2025,Wang:2024,faenzi:2019}. In these structures, a modulated impedance boundary condition (IBC), implemented through subwavelength elements, is used to achieve desired radiation properties. Among other advantages, MTS antennas are generally low profile, lightweight and cost-effective in terms of fabrication. Further, these antennas allow an accurate control of the aperture fields, which enables the generation of pencil, shaped and multiple beams with arbitrary direction and polarization \cite{pandi:2015,Ruiz:2021_RLL,Bodehou:2020,Caminita:2020,Teodorani:2024,Liu:2019_multiBeam,gonzalez:2017}. Multibeam, highly directive MTS antennas are particularly interesting for certain space, military and 5G/6G applications \cite{Guo:2025_overview6G,Ahmed:2023}. Their design presents the challenge of efficiently combining complex feeding networks with high-performance MTS apertures.

Many early and recent modulated MTS antennas integrate a surface wave (SW) launcher at their center, such as SMA connectors \cite{Thanikonda:2025,Teodorani:2024,Wen:2023,Bodehou:2023,faenzi:2019,gonzalez:2017,pandi:2015} or circular waveguides \cite{Pereda:2018,gonzalez:2018}. Cylindrical SW launchers combined with MTSs modulated in a cylindrical coordinate system simplify the generation of circularly-polarized and dual circularly-polarized beams. In addition, such configurations favor compactness. Different center-fed MTSs with multibeam capabilities have been presented \cite{Wang:2024,Wen:2023,Teodorani:2024,Bodehou:2020,faenzi:2019,gonzalez:2017}. However, the small area available for the feeds in the aperture center can limit the number of beams and increase the feeding network complexity. Maximizing the energy coupling between the feed and SW, i.e. feed efficiency, is not straightforward for these designs \cite{Teodorani:2024,Cavillot:2023,minatti:2017}.

Plane-wave excited MTS antennas offer alternative multibeam solutions \cite{Cao:2025,Ruiz:2021_RLL,Ruiz:2021,Yurduseven:2020,wangHybridDigCod:2019,li:2017,Ma:2017}. In these designs, a plane wave illuminates an MTS that is generally modulated along one dimension in a Cartesian coordinate system. Laterally-fed, single-layer designs are simple and can provide a large number of beams as well as wide scanning ranges \cite{li:2017,wangHybridDigCod:2019,Ma:2017}. However, they usually have a large footprint and the aperture efficiency significantly degrades when scanning. Alternatively, multilayer multibeam MTS antennas ensure compactness, reduce scanning loss and enhance feeding efficiency \cite{Cao:2025,Ruiz:2021_RLL,Ruiz:2021,Yurduseven:2020}. Compared to cylindrical SW launchers, plane-wave launchers are generally more complex and their integration into compact structures is challenging. Moreover, a fundamental problem arises in the design of plane-wave excited multibeam MTSs. This problem consists in relating the multiple beam pointing angles with the impinging plane SW direction, and further with the primary feeds' position.

Regarding IBC modulation, sinusoidally modulated MTSs form a category widely exploited in many of the aforementioned antenna designs \cite{Ruiz:2021_RLL,Ruiz:2021,faenzi:2019,gonzalez:2017,Thanikonda:2025,Pereda:2018,gonzalez:2018,Caminita:2020}. When exciting a TM surface wave on a sinusoidally modulated MTS, the periodicity allows to represent the electromagnetic (EM) fields as a sum of infinite spatial harmonics \cite{oliner:59}. For certain harmonics, which wavenumber is smaller than that in the free-space, the SW is progressively transformed into a leaky wave (LW). In other words, it is possible to generate a directive beam from the aperture fields based on the MTS modulation. The direction, polarization and shape of the beam are controlled through the modulation parameters (period, average impedance and modulation amplitude) and nature of the IBC (scalar or tensor). A major advantage of these MTSs is the analytical solution of their dispersion characteristic \cite{oliner:59,caminita:2014}. 

\begin{figure}[t]
	\centering
	\includegraphics[width=1.0\columnwidth]{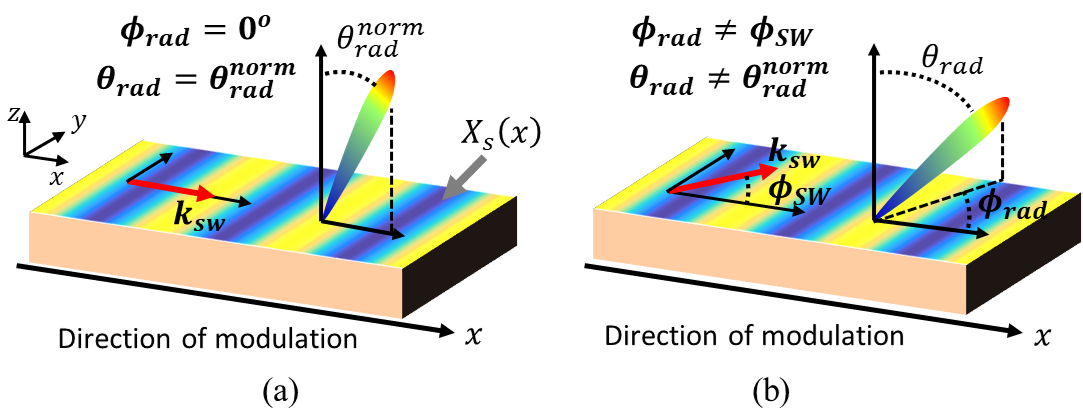}
	\caption{Representation of a plane SW propagating on a 2D MTS sinusoidally modulated along a single direction in a Cartesian coordinate system. Propagation (a) along the modulation direction ($\phi_{SW}=0^\circ$) and (b) at an oblique angle to the modulation direction ($\phi_{SW} \neq 0^\circ$).
	}
	\label{fig:2_obli_conceptFig}
\end{figure}

The canonical 1D problem for sinusoidally modulated MTSs is represented in Fig.~\ref{fig:2_obli_conceptFig}(a). It consists in a plane SW that propagates in the modulation direction of an MTS with spatial sinusoidal variation along one dimension. This problem was solved in the 50s \cite{oliner:59} and its solution is still widely exploited today. An extension to 2D MTSs has significant practical value for designing multibeam antennas. Let us consider 2D MTSs modulated along a single direction in a Cartesian coordinate system \cite{Ruiz:2021_RLL,Ruiz:2021,li:2017,wangHybridDigCod:2019}. If a plane SW is excited along the modulation direction (see Fig.~\ref{fig:2_obli_conceptFig}(a)), the beam direction can only be controlled in elevation. Nevertheless, an additional degree of freedom to control the azimuthal beam pointing angle can be obtained by changing the SW propagation direction, as shown in Fig.~\ref{fig:2_obli_conceptFig}(b). Under this condition, both elevation and azimuth pointing angles are controlled through the SW propagation angle. We note that: i$)$ the azimuthal beam pointing angle does not correspond to the SW propagation angle; ii$)$ the elevation pointing angle does not correspond to that obtained from the 1D canonical problem. As it will be shown, the reason is that for sinusoidally modulated MTSs the radiative spatial harmonic is not generally the fundamental one. Therefore, solving the 2D problem illustrated in Fig.~\ref{fig:2_obli_conceptFig}(b) introduces a procedure for designing multibeam MTS antennas with 2D scanning capability.

Excitation by oblique guided plane waves has been exploited to control the beam pointing angle of slot array antennas in both azimuth and elevation \cite{Bhattacharyya:2015_slotArray}. As for MTSs, expressions have been derived to estimate the beam pointing angles when a plane SW propagates obliquely on an impedance hologram \cite{li:2017}. However, practical conditions such as propagation through different media and related refraction effects have not been considered. Authors in \cite{hwang:1999} provide an exhaustive dispersion analysis of 2D periodic IBCs for TE- and TM-polarized plane waves. Nevertheless, their analysis is focused on waveguiding and does not establish a relation between SW and radiated beam directions. 

In this paper, we provide a general formulation to calculate the beam pointing angles for any SW propagation angle and any spatial harmonic in sinusoidally modulated MTSs. In most practical designs, the plane SW launcher is external to the MTS. Therefore, we also address the problem for discontinuous media. It is shown that Snell's law is fully valid in this case. First, the proposed formulation is numerically validated using several antenna designs with different plane-wave launchers and form factors (both rectangular and circular). Then, we design and fabricate a multibeam modulated MTS antenna excited by a pillbox quasi-optical beamformer. Pillbox beamformers are double-layer structures that favor compactness and can integrate multiple primary feeds, enabling multibeam capabilities \cite{Jasim:2026,Ruiz:2021,Yurduseven:2020,Tekkouk:2015cross}. For this and other practical cases, we use the derived formulation to provide a relation between the beam pointing angles and primary feed positions. We also introduce a matching transition needed to integrate the modulated MTS into the pillbox upper layer. The antenna is fabricated using commercial multilayer PCB technology, resulting in a low-profile, highly directive multibeam prototype with rectangular form factor. Experimental results show the validity of the proposed formulation. 

The article is organized as follows. Section~\ref{sec:obliqueIncidence} addresses the problem of a plane SW that propagates obliquely to the modulation direction on a sinusoidally modulated MTS. The analysis is extended to discontinuous media. Then, the analysis is validated in Section~\ref{sec:additional_validation_examples}, which includes different design examples. The derived formulation is then used in Section~\ref{sec:pillboxMTS_LPantenna_design} to design a compact linearly-polarized multibeam MTS antenna integrating a quasi-optical plane-wave launcher. Experimental results for the MTS antenna are provided in Section~\ref{sec:pillboxMTS_LP_simu}. Conclusions are drawn in Section~\ref{sec:conclusion}.

\section{Oblique Propagation of Plane Surface Waves on Sinusoidally Modulated Metasurfaces}\label{sec:obliqueIncidence}
Fig.~\ref{fig:2_obli_conceptFig} represents a plane SW that propagates along and obliquely to the modulation direction of a sinusoidally modulated 2D MTS. The MTS lies on a grounded dielectric substrate. To characterize the SW direction, we define the angle $\phi_{SW}$ formed by the SW wavevector and direction of modulation. In the following, a direct relation is established between the propagation angle $\phi_{SW}$ and the pointing angles $(\theta_{rad},\phi_{rad})$ of a LW radiated by the MTS. Two main cases are examined. First, a plane SW propagates in a 2D MTS that is sinusoidally modulated along a single direction in a Cartesian coordinate system. Hereinafter, we will also refer to this type of MTSs as unidirectionally modulated. Second, a TEM plane wave guided by a uniform parallel-plate waveguide (PPW) is transmitted to a 2D MTS with unidirectional modulation. In both cases, the 2D MTS is modeled by a modulated tensor sheet impedance $\boldsymbol{\underline{\underline{Z}}_s}=j\boldsymbol{\underline{\underline{X}}_s}$. MTSs modeled by tensor sheet impedances are also denoted as transparent or penetrable \cite{patel:2013april,minatti:2018book}. A time-harmonic dependence of the form $e^{j \omega t}$, where $\omega$ is the angular frequency, is assumed and suppressed throughout the paper.

\subsection{Oblique Propagation on Unidirectionally Modulated Metasurfaces}\label{sec:obliqueInc_SMRS}
\begin{figure}[t]
	\centering
	\includegraphics[width=1.0\columnwidth]{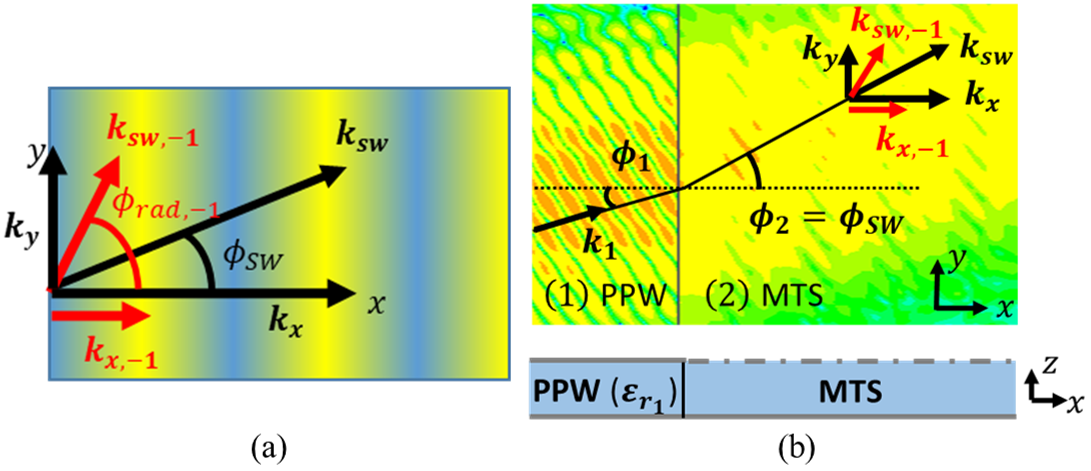}
	\caption{Wavevector analysis of a plane SW with oblique propagation on a unidirectionally sinusoidally modulated MTS. (a) Representation of fundamental ($n=0$) and radiative ($n=-1$) wavevectors. (b) Top ($xy$-plane, upper part) and cross-sectional ($xz$-plane, lower part) view for the case of oblique propagation with change of medium. The top view shows the E-field of a plane wave in a PPW impinging on a modulated MTS. The refraction can be clearly observed.}
	\label{fig:2_obli_schemaSMRS_obliqueInc}
\end{figure}

Let us consider a plane SW that propagates along the modulation direction ($\phi_{SW}=0^\circ$) on a transparent MTS, as shown in Fig.~\ref{fig:2_obli_conceptFig}(a). The MTS is modulated along the $x$-direction and designed to radiate a beam at $\phi_{rad}=0^\circ$ in azimuth and $\theta_{rad}=\theta_{rad}^{norm}$ in elevation. This is done by fixing the period $p$, average reactance $X_{s}^{av}$ and modulation index $M$ of the sheet reactance $\boldsymbol{\underline{\underline{X}}_s}$, which has the form,
\begin{align}
	\begin{split}
		\ & \\
		\boldsymbol{\underline{\underline{X}}_s}=\begin{bmatrix}
			X_s(x)&0\\0&0\\
		\end{bmatrix};\;\ & X_s(x) = X_{s}^{av} [ 1+M\cos(2\pi x/p) ].\\ 
		\ & 
		\label{eq:2_oli_tensorX}
	\end{split}
\end{align}
Now, we consider a plane SW that propagates with an angle $\phi_{SW}\neq0^\circ$ with respect to the MTS modulation direction, as shown in Fig.~\ref{fig:2_obli_conceptFig}(b). In this case, the radiated beam will point at $\phi_{rad} \neq 0^\circ$ and $\theta_{rad}\neq\theta_{rad}^{norm}$.
As illustrated in Fig.~\ref{fig:2_obli_schemaSMRS_obliqueInc}(a), the SW wavevector can be defined as,
\begin{equation}
	\pmb{k_{sw}}=k_x\boldsymbol{\hat{x}}+k_y\boldsymbol{\hat{y}}.
	\label{eq:2obli_kSW}
\end{equation}
From the geometry of the problem, we can readily write the components of $\pmb{k_{sw}}$ as,
\begin{equation}
	k_x=k_{sw} \cos(\phi_{SW}), \hspace{0.5cm}  k_y=k_{sw} \sin(\phi_{SW}).
	\label{eq:2obli_kx_and_ky}
\end{equation}
The periodicity of the sheet impedance enables the expansion of the EM fields just above the surface into Floquet harmonics \cite{oliner:59}. Since the modulation is $x$-directed, spatial harmonics will appear only along $x$, which allows to write,
\begin{equation}
	k_{x,n}=k_x+2\pi n/p,
	\label{eq:2obli_kx}
\end{equation}
where $n$ is the Floquet harmonic index. Therefore, the SW wavenumber for the $n$-th harmonic reads as,
\begin{equation}
	k_{sw,n}=\sqrt{k_{x,n}^2+k_y^2}.
	\label{eq:2obli_kSW_n}
\end{equation}
We can define the fundamental SW wavenumber as $k_{sw}=k_{sw,0}$ and the modal wavenumber along $z$ as,
\begin{equation}
	k_{z,n}=\sqrt{k_0^2-k_{sw,n}^2},
	\label{eq:2obli_kz_n}
\end{equation}
where $k_0$ is the free-space wavenumber. When the radiation condition $k_{sw,n} \leq k_0$ is satisfied, the direction of the beam radiated by the $n$-th spatial harmonic is given by,
\begin{align}
	\theta_{rad,n} &= \arcsin( k_{sw,n}/k_0 ), \label{eq:2obli_theta_rad} \\
	\phi_{rad,n} &= \arctan(k_y/k_{x,n}). \label{eq:2obli_phi_rad}
\end{align}
For non-radiating harmonics, i.e. when $k_{sw,n} > k_0$, expression \eqref{eq:2obli_phi_rad} simply yields the azimuthal propagation angle of the $n$-th SW mode, $\phi_{SW,n}$. Assuming that the MTS is designed to only radiate the harmonic $n=-1$, we can use \eqref{eq:2obli_kx_and_ky} to particularize \eqref{eq:2obli_theta_rad} and \eqref{eq:2obli_phi_rad} as,
\begin{multline}
	\theta_{rad,-1} = \\
	\arcsin \left( \frac{k_{sw}}{k_0} \sqrt{\sin^2(\phi_{SW}) + \left(\cos(\phi_{SW})-\frac{2\pi}{k_{sw} p} \right)^2} \right), 
	\label{eq:2obli_theta_rad-1}
\end{multline}
\begin{align}
	\phi_{rad,-1} &= \arctan \left( \frac{k_{sw}\sin(\phi_{SW})}{k_{sw}\cos(\phi_{SW})-2\pi/p} \right).  
	\label{eq:2obli_phi_rad-1}
\end{align}
For $\phi_{SW}=0^\circ$, one can verify that \eqref{eq:2obli_theta_rad-1} reduces to,
\begin{equation}
	\theta_{rad,-1}=\arcsin(k_{sw}/k_0-2\pi/(k_0 p)),
\end{equation}
as in expression (9) in \cite{patel:2011}, with $\phi_{rad,-1}=0^\circ$. 

The formulation above provides the beam pointing angles of a LW generated by the interaction between SW and MTS (see Fig.~\ref{fig:2_obli_conceptFig}). To obtain $k_{sw}$ used in equations \eqref{eq:2obli_theta_rad} to \eqref{eq:2obli_phi_rad-1}, we need to determine the dispersion characteristic of $k_x$. To this end, we solve the eigen-mode problem of a sinusoidally modulated sheet impedance on a grounded dielectric substrate \cite{caminita:2014,minatti:2018book}. After determining $k_x$, which has the complex form $k_x=\beta_x - j \alpha_x$, we only retain the real part $\beta_x$. We note that the imaginary part $\alpha_x$ has no impact on the radiated beam direction. We also note that $k_x$ does not depend on $\phi_{SW}$, but on the MTS parameters involved in expression \eqref{eq:2_oli_tensorX}.

\subsection{Oblique Propagation With Change of Medium}\label{sec:obliqueInc_PPWtoSMRS}
Next, we consider the scenario shown in Fig.~\ref{fig:2_obli_schemaSMRS_obliqueInc}(b). A TEM plane wave propagates with an angle $\phi_{1}$ with respect to the $x$-direction in a lossless dielectric-filled PPW with relative permittivity $\varepsilon_{r_1}$ (medium 1). The PPW lies in the $xy$-plane. The plane wave is then transmitted to a 2D MTS sinusoidally modulated along $x$ (medium 2), as the MTS described in Section~\ref{sec:obliqueInc_SMRS}. The field distribution for the TEM plane wave in the PPW can be written as \cite{pozar:2012},
\begin{align}
	\boldsymbol{E}(x,y)&=-\boldsymbol{\hat{z}} \frac{V_0}{h}e^{-j \boldsymbol{k_1} \cdot \left(\boldsymbol{\hat{x}}x+\boldsymbol{\hat{y}}y\right)},\\
	\boldsymbol{H}(x,y)&=-(\boldsymbol{\widehat{k}_1} \times \boldsymbol{\hat{z}}) \frac{V_0}{\eta_1 h}e^{-j\boldsymbol{k_1} \cdot \left(\boldsymbol{\hat{x}}x+\boldsymbol{\hat{y}}y\right)},
	\label{eq:2obli_EH_PPW}
\end{align}
where $V_0$ is the potential between the PPW plates, $d$ is the thickness of the PPW and $\eta_1$ is the wave impedance in the PPW. The wavevector in the PPW has a direction $\boldsymbol{\widehat{k}_1}$ and can be expressed as,
\begin{equation}
	\boldsymbol{k_1}=k_1( \boldsymbol{\hat{x}}\cos\phi_1 + \boldsymbol{\hat{y}}\sin\phi_1 ),
	\label{eq:2obli_k1vector}
\end{equation}
where $k_1=k_0\sqrt{\varepsilon_{r_1}}$ is the wavenumber in the PPW. Using \eqref{eq:2obli_k1vector} in \eqref{eq:2obli_EH_PPW}, the phase of the magnetic field $\boldsymbol{H}$ can be written as,
\begin{equation}
	\Psi_1 = e^{-j \left( k_1\cos\phi_1 \cdot x + k_1\sin\phi_1 \cdot y \right)}.
	\label{eq:2obli_phase_PPW}
\end{equation}
Since the dielectric-filled PPW is assumed to be lossless, $k_1$ is purely real. In the MTS, the phase of the magnetic field can be written as \cite{oliner:59},
\begin{equation}
	\Psi_2 = e^{ -jk_yy } \sum_{n=-\infty}^{\infty} e^{ -jk_{x_n}x } e^{ -j k_{z_n}z }.
	\label{eq:2obli_phase_SMRS}
\end{equation}
The tangential magnetic field is continuous across the $yz$-plane interface that separates medium 1 and 2 (see Fig.~\ref{fig:2_obli_schemaSMRS_obliqueInc}(b)). This results in the phase-matching condition $\Psi_1=\Psi_2$ at the interface. By using \eqref{eq:2obli_kx_and_ky} in \eqref{eq:2obli_phase_SMRS}, and placing the interface at $x=0$ and the MTS plane at $z=0$, the phase matching condition yields,
\begin{equation}
	k_1\sin\phi_1 = k_{sw}\sin\phi_{SW}.
	\label{eq:2obli_snellLaw}
\end{equation}
Equation \eqref{eq:2obli_snellLaw} is the Snell's law of refraction. For the medium containing the modulated MTS, equation \eqref{eq:2obli_snellLaw} simply involves the SW fundamental harmonic $k_{sw}$. The unknown to solve upon \eqref{eq:2obli_snellLaw} is $\phi_{SW}$. Then, one can just use $\phi_{SW}$ in \eqref{eq:2obli_theta_rad} and \eqref{eq:2obli_phi_rad} to obtain the direction of the beam radiated by any harmonic $n$. In this analysis we have ignored the reflected fields, since they have no impact on the calculation of $\phi_{SW}$. The derived closed-form expressions to calculate radiated beam directions are numerically and experimentally validated in the following sections.

\section{Numerical Validation of Beam Direction Prediction for Plane-Wave Excited Modulated Metasurfaces}\label{sec:additional_validation_examples}
In this section, we demonstrate the validity of the analytical model derived in Section~\ref{sec:obliqueIncidence} for different MTS antenna designs. Simulated (numerical) and theoretical results are presented for various unidirectionally modulated MTSs with different plane-wave launchers. All modulations in this section have the form given in \eqref{eq:2_oli_tensorX}. In addition, for all the presented designs we consider the pointing angles of the beam generated by the spatial Floquet harmonic $n=-1$. The parameters used for the various modulated MTSs in this section are listed in Table~\ref{table:2obli_table_MTSdetails}. Commercial full-wave solvers \cite{HFSS,CST} have been used to obtain the presented numerical results.

\subsection{Modulated Metasurfaces With Embedded Feed}\label{sec:obliqueInc_SMRS_valid}
\begin{figure}[t]
	\centering
	\includegraphics[width=1\linewidth]{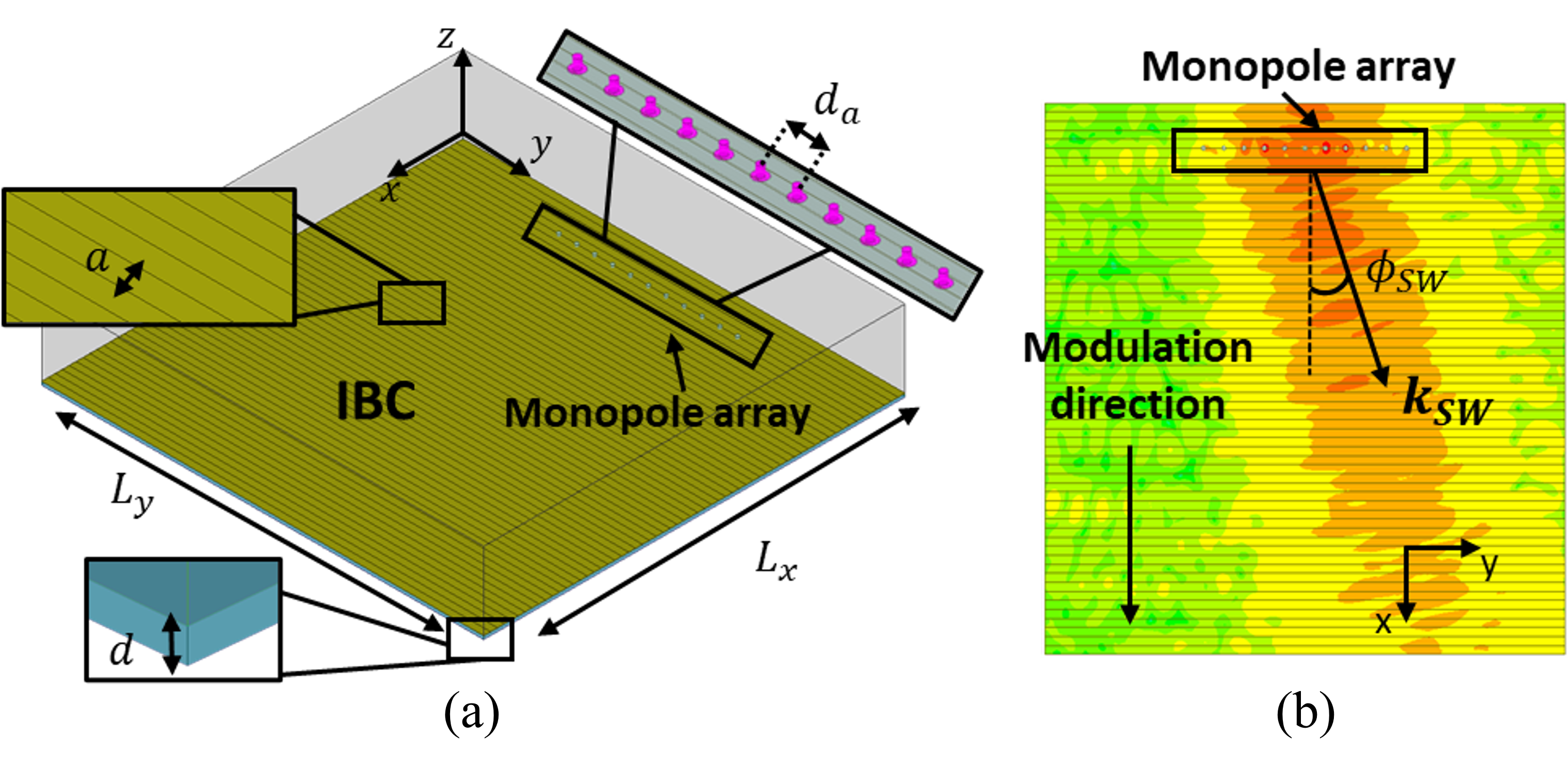}
	\caption{Design of modulated MTS with embedded feed. (a) Modulated MTS formed from ideal IBC strips with an embedded array of monopoles used as plane SW launcher. Array elements are separated by a distance $d_a=\lambda_{sw}/2$ ($\lambda_{sw}=2\pi/k_{sw}$). An air box is placed above the MTS, and its faces are assigned radiation BCs. The MTS dimension along $y$ is $L_y=L_x$ (see Table~\ref{table:2obli_table_MTSdetails}). (b) Top view of modulated MTS with an excited plane SW propagating obliquely to the direction of modulation.}
	\label{fig:2obli_monopMTS_structure_and_schema}
\end{figure}

First, we use the structure shown in Fig.~\ref{fig:2obli_monopMTS_structure_and_schema}(a). In this case, which implements the scenario presented in Fig.~\ref{fig:2_obli_schemaSMRS_obliqueInc}(a), a plane SW is directly generated on the MTS by an embedded array of monopoles (see Fig.~\ref{fig:2obli_monopMTS_structure_and_schema}(b)). The array contains $N=11$ elements that generate a Gaussian amplitude taper of the plane SW. The SW propagation angle $\phi_{SW}$ is changed by imposing a phase shift between the array elements of $\Delta\phi_a=-k_{sw} d_a \sin\phi_{SW}$, where $d_a=\lambda_{sw}/2$ is the distance between elements and $\lambda_{sw}$ is the wavelength on the MTS. The modulated MTS is implemented by parallel strips of width $a=\lambda_0/10$ ($\lambda_0$ is the free-space wavelength) at $10$ GHz (see Fig.~\ref{fig:2obli_monopMTS_structure_and_schema}(a)), each of which is assigned a local constant IBC value. The IBC strips are placed on a grounded dielectric substrate with $\varepsilon_r=6. 15$, $\tan{\delta}=0.002$ and thickness $d$. We limit the SW propagation angle to $\phi_{SW}\leq30^\circ$. For larger values of $\phi_{SW}$, the directions of SW and modulation become too oblique for radiating a proper beam.

\begin{figure}[t]
	\centering
	\includegraphics[width=1\linewidth]{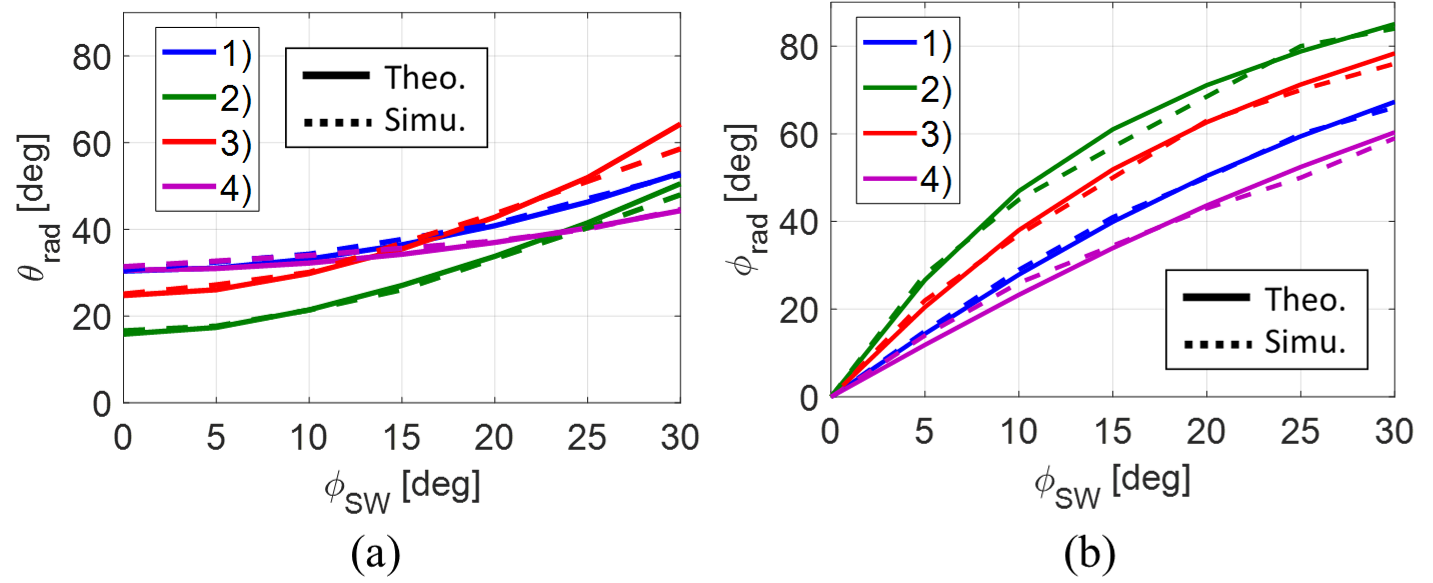}
	\caption{Theoretical and simulated beam pointing angles in (a) $\theta$ and (b) $\phi$ for different designs of modulated MTSs with embedded feed. The pointing angles are given as a function of the SW propagation angle. Design parameters of the various MTSs are given in Table~\ref{table:2obli_table_MTSdetails}.}
	\label{fig:2obli_monopMTS_results}
\end{figure}

The simulated and theoretical beam pointing angles are shown in Fig.~\ref{fig:2obli_monopMTS_results} for several configurations of the modulated MTS. The parameters for the various MTS configurations are provided in Table~\ref{table:2obli_table_MTSdetails}. The theoretical curves are obtained using the SW propagation angle $\phi_{SW}$ in \eqref{eq:2obli_theta_rad-1} and \eqref{eq:2obli_phi_rad-1}. As observed, the agreement between simulated and theoretical results is very good for both $\theta$ and $\phi$ angles. Further, an interesting effect is noticeable concerning the attainable scanning range. As shown in Fig.~\ref{fig:2obli_monopMTS_results}, by tilting the plane SW up to only $30^\circ$ we obtain $\theta > 40^\circ$ and $\phi \geq 60^\circ$ for all designs. In particular, the azimuthal angle spans to almost $90^\circ$ for design 2). This effect in $\phi$ is described in Fig.~\ref{fig:2_obli_schemaSMRS_obliqueInc}(a), where it is shown that $k_{x,-1}<k_x$ while $k_{y,-1}=k_y$ ($k_{y,-1}$ is the $y$-component of $k_{sw,-1}$). This implies that $\phi_{rad,-1} > \phi_{SW}$.

{{\setlength{\tabcolsep}{4pt}
		\begin{table}[t]
			\centering
			\caption{Parameters of modulated MTSs used in design examples of Section~\ref{sec:additional_validation_examples}. Designs 1) to 8) integrate MTSs with transparent/sheet impedance, whereas the MTS in design 9) implements an opaque impedance.}
			\begin{tabular}{ |c|c|c|c|c|c|c|c| }
				\hline
				Design & $\frac{X^{av}}{\eta_0}$ & $M$ & $p$ [mm] & \makecell{$\theta_{rad}^{norm}$ \\ \,[deg]} & $d$ [mm] & \makecell{$f_0$ \\ \,[GHz]} & $\frac{L_x}{\lambda_0}$ \\
				\hline
				$1)$  & $-0.5$ & $0.35$ & $31.0$ & $30$ & $1.905$ & $10$ & 12 \\
				$2)$  & $-0.43$ & $0.4$ & $23.7$ & $15$ & $1.905$ & $10$ & 12 \\
				$3)$  & $-0.28$ & $0.4$ & $22.28$ & $23$ & $1.905$ & $10$ & 12 \\
				$4)$  & $-0.64$ & $0.46$ & $41.5$ & $30$ & $1.575$ & $10$ & 12 \\
				$5)$  & $-0.5$ & $0.43$ & $21.14$ & $12$ & $1.905$ & $10$ & 12 \\
				$6)$  & $-0.86$ & $0.46$ & $26.6$ & $9$ & $1.905$ & $10$ & 12 \\
				$7)$  & $-0.25$ & Tap. & $8.9$ & $-30$ & $0.64$ & $20.7$ & 15 \\
				$8)$  & $-0.25$ & Tap. & $7.4$ & $-10$ & $0.64$ & $20.7$ & 15 \\
				$9)$  & $0.6$ & $0.82$ & $9.45$ & $5$ & $2$ & $30$ & 13 \\
				\hline
			\end{tabular}
			\label{table:2obli_table_MTSdetails}
		\end{table}
	}
	
	\subsection{Modulated Metasurfaces Laterally Fed}\label{sec:obliqueInc_PPWtoSMRS_valid}
	Next, we use the design shown in Fig.~\ref{fig:2obli_monopPPW_structure_Phi1vsPhi2}, which implements the scenario presented in Fig.~\ref{fig:2_obli_schemaSMRS_obliqueInc}(b) and consists of a modulated MTS placed after a PPW. Both the PPW and MTS use the same dielectric as in Section~\ref{sec:obliqueInc_SMRS_valid}, and the operating frequency is $f_0=10$ GHz. The modulated metasurface is implemented using parallel IBC strips of constant width ($a=\lambda_0/10$) and locally assigned impedance values. An array of $N = 21$ monopoles with Gaussian amplitude taper is embedded within the dielectric-filled PPW to excite the structure. The number of elements in the feeding array has been increased to compensate for power loss due to reflections at the PPW--MTS interface. A matching transition is omitted for simplicity. The excited plane-wave propagation angle $\phi_{1}$ is changed by imposing a phase shift between the array elements of $\Delta\phi_a=-k_1 d_a \sin\phi_{1}$, where $d_a=\lambda_1/2$ is the distance between elements and $\lambda_1$ is the wavelength in the PPW (see Fig.~\ref{fig:2obli_monopPPW_structure_Phi1vsPhi2}). This angle is limited to $\phi_1\leq20^\circ$ for the same reasons given in Section~\ref{sec:obliqueInc_SMRS_valid}. In this case, the plane-wave angular limit is lower since the refraction from PPW to MTS yields $\phi_{SW}>\phi_{1}$ (see Fig.~\ref{fig:2_obli_schemaSMRS_obliqueInc}(b)).
	
	\begin{figure}[t]
		\centering
		\includegraphics[width=0.9\linewidth]{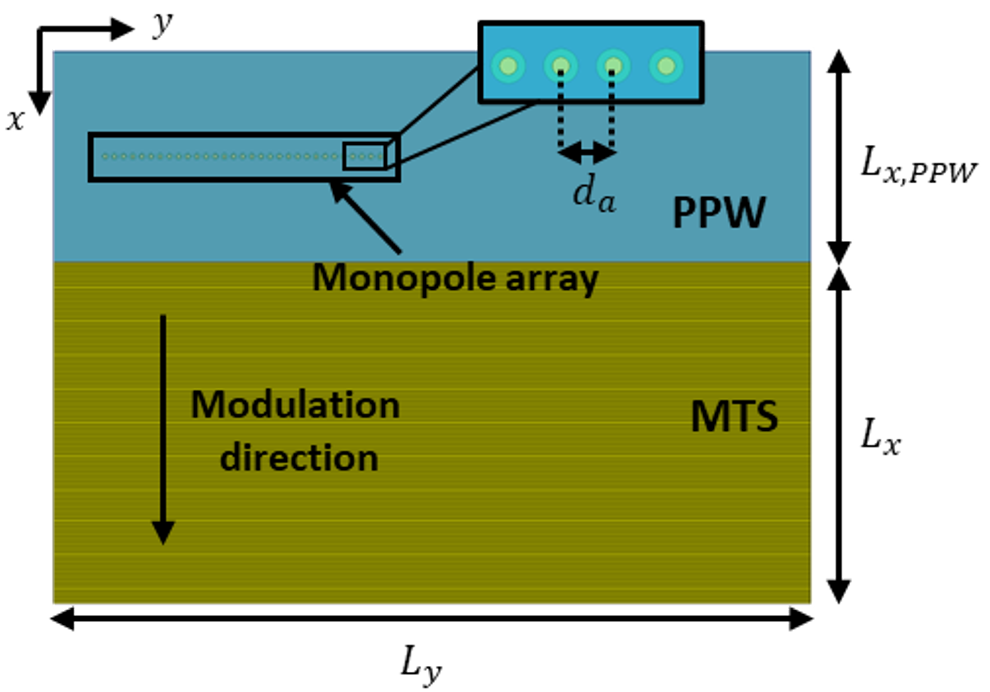}
		\caption{Design of modulated MTS laterally fed. The modulated MTS is formed from ideal IBC strips and excited by an array of monopoles embedded in a dielectric-filled PPW. Array elements are separated by a distance $d_a=\lambda_1/2$ (where $\lambda_1=\lambda_0/\sqrt{\varepsilon_{r}}$ is the wavelength in the PPW). An air box (not shown in the figure) is placed above the MTS, and its faces are assigned radiation BCs. The MTS dimension along $y$ is $L_y=L_x+L_{x,PPW}$ (see Table~\ref{table:2obli_table_MTSdetails}), where $L_{x,PPW}=L_x/3$.}
		\label{fig:2obli_monopPPW_structure_Phi1vsPhi2}
	\end{figure}
	\FloatBarrier
	\begin{figure}[!t]
		\centering
		\includegraphics[width=1\linewidth]{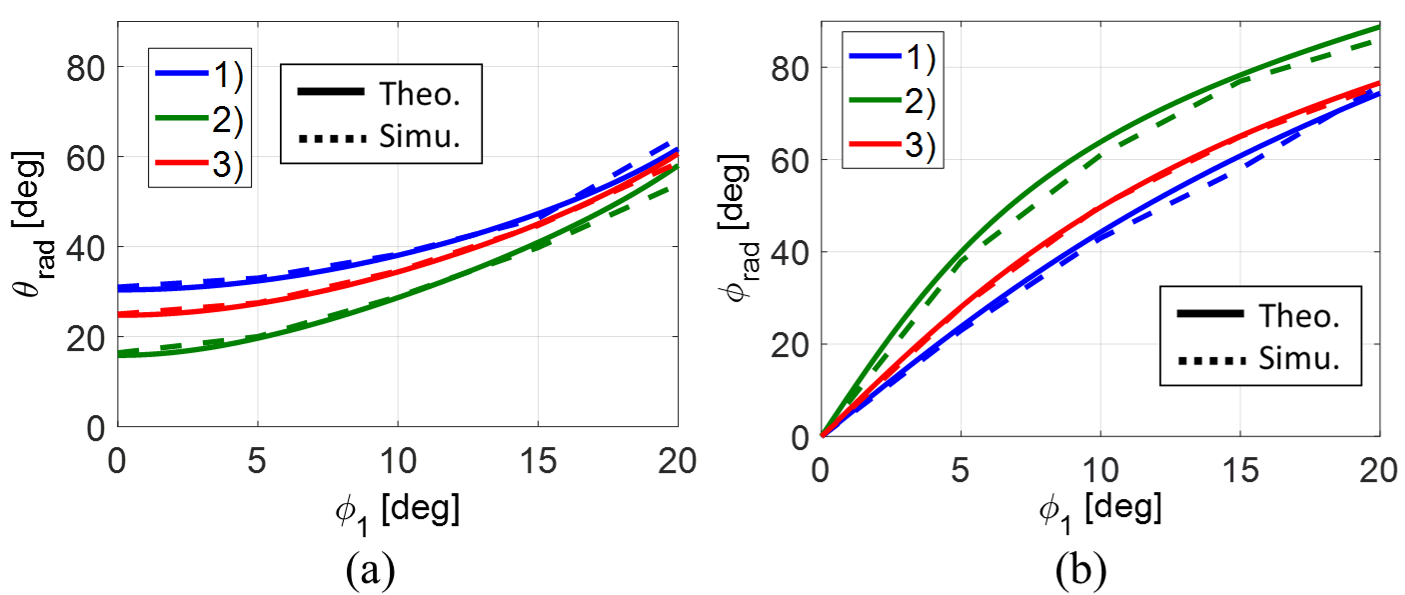}
		\caption{Theoretical and simulated beam pointing angles in (a) $\theta$ and (b) $\phi$ for different designs of modulated MTSs laterally fed. The pointing angles are given as a function of the plane wave angle produced by the launcher. Design parameters of the various MTSs are given in Table~\ref{table:2obli_table_MTSdetails}.}
		\label{fig:2obli_monopPPW_results}
	\end{figure}
	
	Simulated and theoretical results are shown in Fig.~\ref{fig:2obli_monopPPW_results} for different MTS designs. The MTS parameters for each design are listed in Table~\ref{table:2obli_table_MTSdetails}. For these designs, the theoretical curves plotted in Fig.~\ref{fig:2obli_monopPPW_results} are obtained using \eqref{eq:2obli_snellLaw} to calculate $\phi_{SW}$ at first. Then, we use \eqref{eq:2obli_theta_rad-1} and \eqref{eq:2obli_phi_rad-1} to calculate the beam pointing angles. A good agreement is observed between simulated and theoretical results. Despite tilting the plane wave up to only $20^\circ$, these designs provide a very similar scanning range to that in Section~\ref{sec:obliqueInc_SMRS_valid}. This is due to the refraction at the PPW-MTS interface, and at the expense of requiring a matching transition in practice. The design of a matching transition is described later in the paper for the realization of a MTS antenna prototype.

	\subsection{Modulated Metasurfaces With Quasi-Optical Pillbox Feed}\label{sec:obliqueInc_quasiOptPillbox}
	The designs presented in Section~\ref{sec:obliqueInc_SMRS_valid} and \ref{sec:obliqueInc_PPWtoSMRS_valid} have been primarily implemented for validating the formulation introduced in Section~\ref{sec:obliqueIncidence}. In this and the following section, we present quasi-optically fed designs that lead to more practical prototypes. Here, we use pillbox beamformers as plane-wave launchers. A pillbox is a double-layer quasi-optical structure that integrates multiple primary feeds and a parabolic reflector in the lower layer, as shown in Fig.~\ref{fig:3_schemaPillboxMTS}(a). The feeds and reflector are configured to generate multi-directional plane waves in the upper layer. These plane waves can illuminate a modulated MTS and produce multiple beams with different directions \cite{Ruiz:2021}. The working principle of a pillbox-fed modulated MTS is illustrated in Fig.~\ref{fig:3_schemaPillboxMTS}(b). Pillbox beamformers are discussed in detail in Section~\ref{sec:pillboxMTS_LPantenna_design}. 
	
	In this section, two different pillboxes are used, operating respectively at 10 GHz and 20.7 GHz. Their dimensions are listed in Table~\ref{table:3_pillbox_details_validationEx}. For both pillboxes, the primary feed in the bottom layer is progressively shifted along $y$ up to $20$ mm from its central position (see Fig.~\ref{fig:3_schemaPillboxMTS}(a)). Further, the modulated MTS used in the top layer is formed from parallel strips with locally constant IBC values (same as the MTS shown in Fig.~\ref{fig:2obli_monopMTS_structure_and_schema} and Fig.~\ref{fig:2obli_monopPPW_structure_Phi1vsPhi2}). We use the same dielectric in the pillbox and MTS ($\varepsilon_r=6. 15$, $\tan{\delta}=0.002$ and thickness $d$). Fig.~\ref{fig:2obli_pillboxMTS_results} shows the simulated and theoretical beam pointing angles for several pillbox-fed MTS designs. The theoretical curves are obtained as follows. First, we establish a relation between the feed position and the azimuthal angle $\phi_{1}$ of the generated plane wave in the pillbox top layer. To establish this relation, we use a simple paraxial approximation given by $\phi_1=\arctan(y_{port}/F_d)$, where $y_{port}$ is the feed center position with respect to the parabola focal point, placed at $y=0$, and $F_d$ is the focal length (see Fig.~\ref{fig:3_schemaPillboxMTS}(a)). Then, we use \eqref{eq:2obli_snellLaw} to calculate $\phi_{SW}$, and \eqref{eq:2obli_theta_rad-1} and \eqref{eq:2obli_phi_rad-1} to obtain the beam pointing angles. The design parameters for the various MTSs are given in Table~\ref{table:2obli_table_MTSdetails}. MTS designs 7) and 8) have a tapering profile of the modulation index $M$. This format of $M$ is thoroughly described later in Section~\ref{sec:MTS_design}. As observed in Fig.~\ref{fig:2obli_pillboxMTS_results}, simulated and theoretical results are in very good agreement. We note that by shifting the pillbox primary feed a few millimeters (up to $20$ mm), a large scanning range is achievable. In particular, design 7) and 8) provide a variation of the beam pointing angle of $\Delta\theta\approx40^\circ$, $\Delta\phi_{7)}\approx45^\circ$ and $\Delta\phi_{8)}\approx60^\circ$. Therefore, primary feeds should be precisely positioned in these systems, since a lateral deviation of a few millimeters can lead to a non-negligible variation of the beam direction.
	
	\begin{figure}[t]
		\centering
		\includegraphics[width=1\columnwidth]{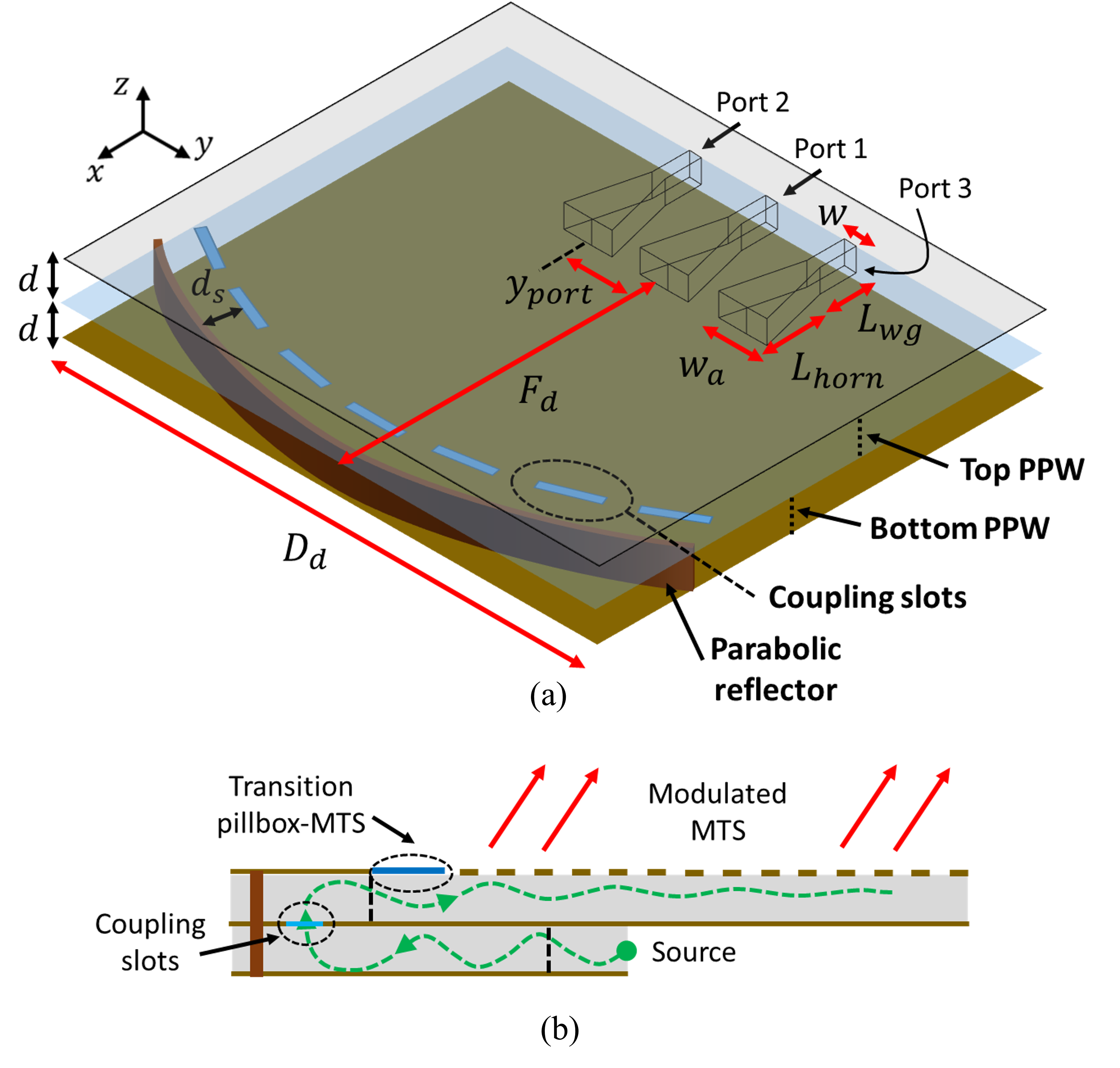}
		\caption{Pillbox quasi-optical beamformer. (a) Details of a pillbox used to generate plane waves with different directions. (b) Schematic section view showing the working principle of a pillbox-fed MTS antenna. This working principle is generally valid for MTS antennas fed by double-layer quasi-optical beamformers. }
		\label{fig:3_schemaPillboxMTS}
	\end{figure}
	
	\begin{figure}[t]
		\centering
		\includegraphics[width=1\linewidth]{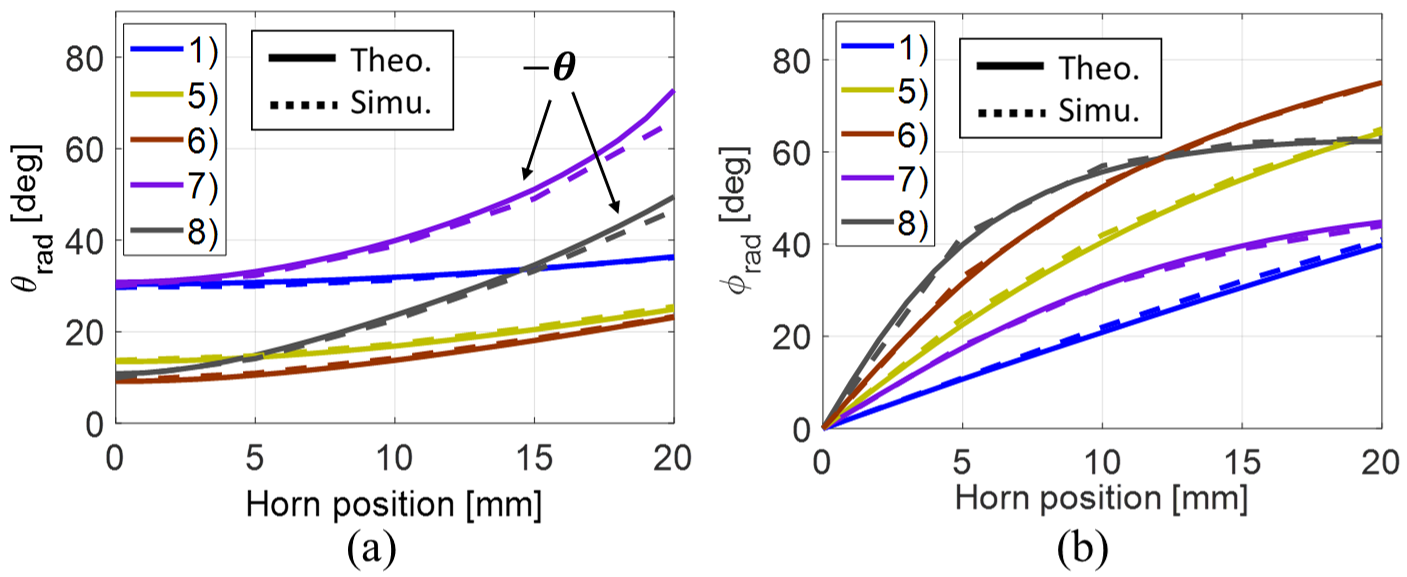}
		\caption{Theoretical and simulated beam pointing angles in (a) $\theta$ and (b) $\phi$ for different designs of pillbox-fed modulated MTSs. The pointing angles are given as a function of the primary feed position. Design parameters of the various MTSs are given in Table~\ref{table:2obli_table_MTSdetails}. In (a), elevation angles for design 7) and 8) have negative values.}
		\label{fig:2obli_pillboxMTS_results}
	\end{figure}
	
	{\setlength{\tabcolsep}{4pt}
		\begin{table}[t]
			\centering
			\caption{Operating frequency and parameters in millimeters for the pillbox beamformers used in this work.}
			\begin{tabular}{ |c|c|c|c|c|c|c|c|c|c|c| }
				\hline
				$f_0$ & $d$ & $F_d$ & $D_d$ & $w$ & $w_a$ & $y_{port}$ & $L_{wg}$ & $L_{horn}$ \\
				\hline
				$10$ GHz & $1.905$ & $109$ & $278$ & $8$ & $11$ & $[0,20]$ & $5$ & $18$ \\
				\hline
				$20.7$ GHz & $1.28$ & $101$ & $252$ & $5$ & $6$ & $[0,20]$ & $2$ & $7.1$ \\
				\hline
			\end{tabular}
			\label{table:3_pillbox_details_validationEx}
		\end{table}
	}
	
	\begin{figure}[t]
		\centering
		\includegraphics[width=1\columnwidth]{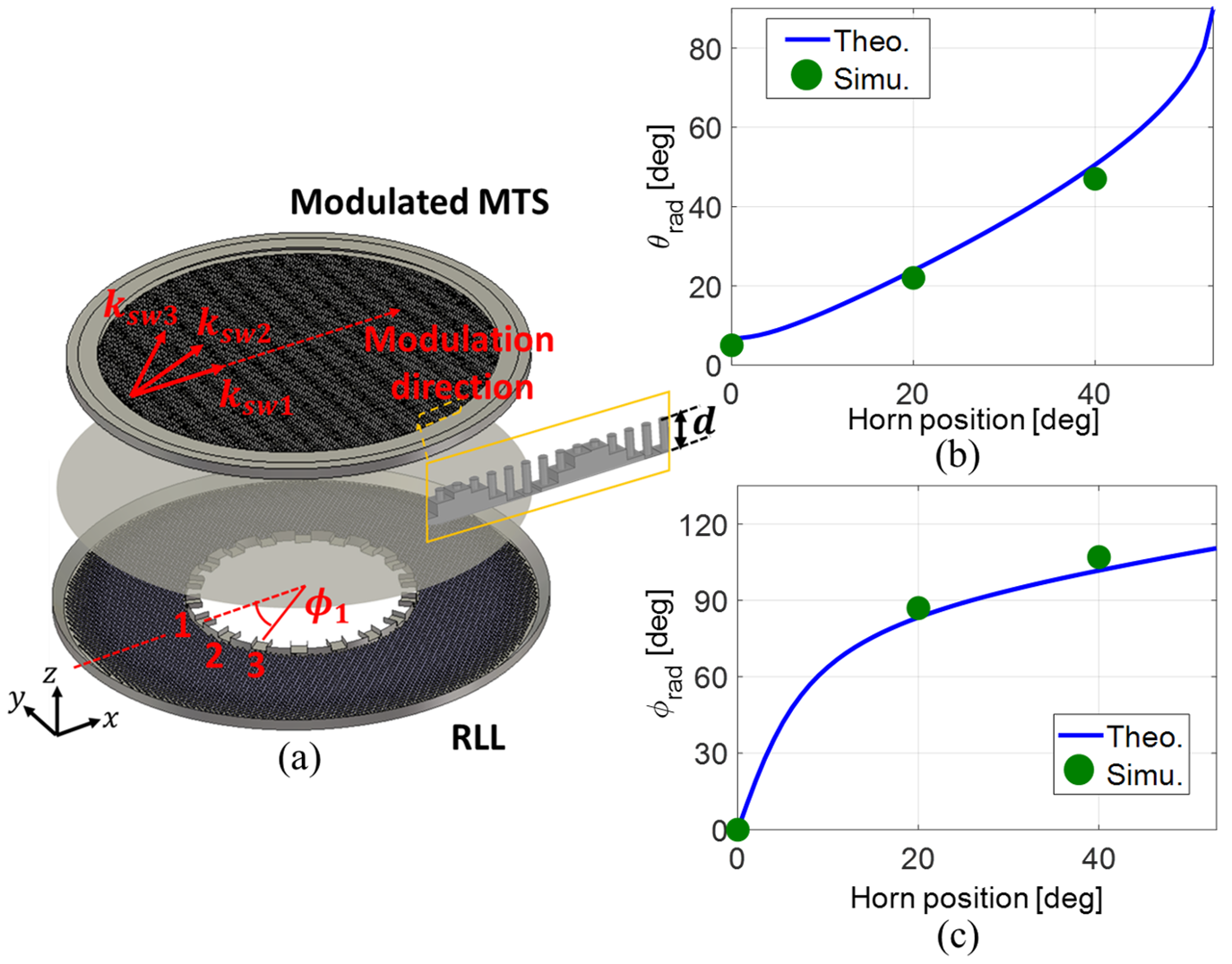}
		\caption{Design of modulated MTS quasi-optically fed by a reflecting Luneburg lens (RLL). (a) Modulated MTS implemented by metallic pins (upper layer) and excited by an azimuthally-symmetric plane-wave launcher (lower layer). Theoretical and simulated beam pointing angles in (a) $\theta$ and (b) $\phi$. The pointing angles are given as a function of the feed angular position. Design parameters of the modulated MTS correspond to design 9) in Table~\ref{table:2obli_table_MTSdetails}.}
		\label{fig:RLL_MTS}
	\end{figure}

	\subsection{Modulated Metasurface With Reflecting Luneburg Lens}\label{sec:obliqueInc_quasiOpt}
	The final design example corresponds to the MTS antenna presented in \cite[Section~IV-D]{Ruiz:2021_RLL} and shown in Fig.~\ref{fig:RLL_MTS}(a). In this antenna, a cylindrical wavefront launched by a primary feed in the lower layer is collimated and coupled to the upper layer by a circular corner reflector. The structure is azimuthally symmetric, able to produce a plane wave in the upper layer whose direction depends on the active primary feed's angular position. By placing a modulated MTS in the upper layer, highly directive beams can be radiated through the working principle described in Fig.~\ref{fig:3_schemaPillboxMTS}(b). For the antenna shown in Fig.~\ref{fig:RLL_MTS}(a), the angle $\phi_{1}$ of the excited plane wave corresponds directly to the angular position of the active feed in the lower layer. Then, the beam pointing angles are obtained by first using \eqref{eq:2obli_snellLaw} to calculate $\phi_{SW}$, and then \eqref{eq:2obli_theta_rad-1} and \eqref{eq:2obli_phi_rad-1}. The modulated MTS used in this design corresponds to design 9) in Table~\ref{table:2obli_table_MTSdetails}, which operates at 30 GHz. The MTS is implemented using metallic pins with uniform height and varying ground level (see inset in Fig.~\ref{fig:RLL_MTS}(a)). For this all-metal MTS design, parameter $d$ in Table~\ref{table:2obli_table_MTSdetails} denotes the height of the pins. Simulated and theoretical results corresponding to this design are shown in Fig.~\ref{fig:RLL_MTS}(b)-(c). Due to spatial constraints in the focal region of the lower layer (see Fig.~\ref{fig:RLL_MTS}(a)), the angular separation between adjacent feeds is $20^\circ$. Up to three different feeds are excited in the simulated design (placed at $\phi_{1}=0^\circ, 20^\circ, 40^\circ$). Agreement between simulated and theoretical results is very good for this structure as well, which demonstrates the generality of the proposed formulation.

	\section{Design of a Quasi-Optically-Fed Multibeam Metasurface Antenna}\label{sec:pillboxMTS_LPantenna_design}
	The analysis presented in Section~\ref{sec:obliqueIncidence} provides a framework for designing quasi-optically-fed MTS antennas with multibeam capabilities. In this section, we introduce a linearly-polarized modulated MTS antenna fed by a pillbox beamformer. First, we describe the quasi-optical system. Then, the design of the MTS and its implementation by subwavelength unit cells are presented. The antenna operates in the 19.7-21.7 GHz band, with a central frequency $f_0=20.7$ GHz.

	\subsection{Pillbox Quasi-Optical Beamformers}\label{sec:pillbox}
	A pillbox quasi-optical beamformer is a double-layer system that consists of a 180$^\circ$ PPW bend with parabolic profile \cite{Rotman:1958pillbox}, as shown in Fig.~\ref{fig:3_schemaPillboxMTS}(a). This structure collimates a cylindrical wavefront launched by a primary feed in the bottom layer. Among other advantages, pillbox beamformers lead to compact antenna designs with discrete beam scanning in azimuth and elevation \cite{Tekkouk:2015cross}. The beam scanning is enabled by arranging several feeds at the parabola's focal plane in the bottom layer (see Fig.~\ref{fig:3_schemaPillboxMTS}(a)). Indeed, the lateral displacement of the primary feed will change the generated plane-wave wavevector direction in the top layer. Some antenna designs use the multi-directional plane waves generated by the pillbox to illuminate a radiating aperture in the top layer \cite{Jasim:2026,Yurduseven:2020,Tekkouk:2015cross}. In our case, we use a unidirectionally modulated MTS as radiating aperture \cite{Ruiz:2021,Ruiz:2020_eucap,Bertrand:2024}. The working principle of the pillbox-fed MTS antenna is illustrated in Fig.~\ref{fig:3_schemaPillboxMTS}(b).
	
	In order to characterize the beam radiated by the pillbox-fed MTS antenna, let us consider the scenarios shown in Fig.~\ref{fig:3_Hplane_Eplane_definition}. In Fig.~\ref{fig:3_Hplane_Eplane_definition}(a), we assume that the central feed of the pillbox, placed at the parabola's focal point, is active. In the MTS at the top layer, this produces a plane SW along the $x$-direction with $\phi_{SW}=0^\circ$. For this configuration, the E-plane of the radiated beam corresponds to the $xz$-plane. In Fig.~\ref{fig:3_Hplane_Eplane_definition}(b), we consider that an offset feed of the pillbox is active, which leads to a plane SW on the metasurface with $\phi_{SW} \ne 0^\circ$. The E-plane in this case is orthogonal to the $xy$-plane and contains the origin and the direction of maximum directivity. 
	In both cases shown in Fig.~\ref{fig:3_Hplane_Eplane_definition}, the H-plane of the beam is orthogonal to the E-plane and also contains the origin and the point of maximum directivity. While the beam properties in the E-plane are mainly dictated by the modulated MTS, the beam properties in the H-plane are mainly dictated by the pillbox beamformer.
	
	\begin{figure}[t]
		\centering
		\subfloat[]{
			\centering
			\includegraphics[width=0.5\columnwidth]{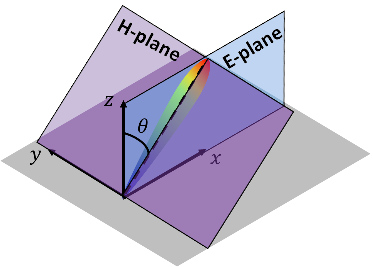}
			\label{fig:3_Hplane_Eplane_definition_norm}
		}
		\subfloat[]{
			\centering
			\includegraphics[width=0.5\columnwidth]{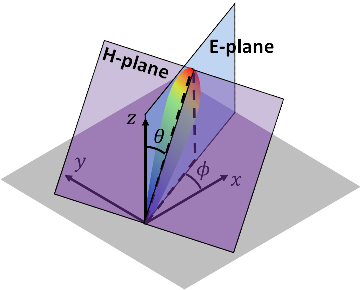}
			\label{fig:3_Hplane_Eplane_definition_offset}
		}
		\caption{Definition of E-plane (blue sheet) and H-plane (purple sheet) for the proposed pillbox-fed MTS antenna. The antenna, represented here by the gray sheet, lies in the $xy$-plane. Radiated beam produced by (a) the pillbox central feed and (b) a pillbox laterally-displaced (offset) feed.}
		\label{fig:3_Hplane_Eplane_definition}
	\end{figure}
	
	In the proposed antenna design, we intend to equalize the beamwidth in the H-plane with that in the E-plane to produce a symmetric pencil beam at $f_0$. To this end, we use the pillbox with $f_0=20.7$ GHz listed in Table~\ref{table:3_pillbox_details_validationEx}. The primary feeds of the pillbox consist of H-plane horns with rectangular waveguide inputs of length $L_{wg}$ and width $w$, as illustrated in Fig.~\ref{fig:3_schemaPillboxMTS}(a). The width $w$ is selected to guarantee uni-modal excitation of the fundamental TE$_{10}$ mode. Our antenna integrates in both layers a Rogers RO3006 substrate, with $\varepsilon_r=6.15$, $\tan\delta=0.002$ and thickness $d=1.28$ mm. At the operating band, the substrate produces non-negligible propagation losses. To minimize these losses, the pillbox focal distance is shortened while ensuring a symmetric beam. This shortening of the pillbox bottom layer is illustrated in Fig.~\ref{fig:3_schemaPillboxMTS}(b). As for the rest of parameters listed in Table~\ref{table:3_pillbox_details_validationEx}, $d$ is the thickness of each layer, $F_d$ is the focal distance and $D_d$ is the diameter of the parabolic reflector.

	\subsection{Design of Modulated Metasurface}\label{sec:MTS_design}
	We describe first the ideal and required profile of the modulated sheet reactance $X_s(x)$. When the illuminating SW propagates along the modulation direction ($\phi_{SW}=0^\circ$, see Fig.~\ref{fig:2_obli_conceptFig}(a)), the elevation pointing angle is fixed to $\theta_{rad}^{norm}=-12^\circ$ at $f_0=20.7$ GHz. This configuration ensures that only the spatial harmonic $n=-1$ radiates. In addition, the SW power ratio is enforced to be $P_L⁄P_0 =-12$ dB over the entire frequency band, where $P_L$ and $P_0$ are the SW powers at the end and input of the MTS, respectively. The length of the modulated MTS along $x$ is $L_x=14.5\lambda_0$. To obtain the required $\theta_{rad}^{norm}$ and SW power ratio $P_L⁄P_0$ as well as to maximize the radiation efficiency, we choose $X_{s}^{av}=-0.83\eta_0$, $p=7.6$ mm and the tapered modulation index $M(x)$ shown in Fig.~\ref{fig:3_tapProf_Xs_g_LP}(a). The spatially dependent modulation index is obtained as follows. We first impose an aperture power density profile $S(x)$ that follows the normalized expression \cite{minatti:2017},
	\begin{equation}
		\frac{S(x)}{S_{max}} = 
		\begin{cases} 
			1, & 0 \leq x < 1-r/2. \\ 
			\frac{1}{2} [1+\cos(\frac{2\pi}{r}[x-1+r/2])], & 1-r/2 < x \leq 1. \\
		\end{cases}
		\label{eq:2_tukeyAggress}
	\end{equation}
	The profile of $S(x)$ is defined over a normalized length. From the imposed aperture power density, the corresponding leaking factor is calculated as \cite{volakis:2007ch11},
	\begin{equation}
		\alpha(x) = \frac{S(x)/2}
		{ \dfrac{1}{\varepsilon_{conv}} \integSxLDen - \integSxDen }.
		\label{eq:2_alpha_x_total}
	\end{equation}
	The tapering of $\alpha$ allows to maximize the aperture efficiency \cite{minatti:2017,Bodehou:2020_effMTS}. For a damping factor $r=0.32$ in \eqref{eq:2_tukeyAggress} and a conversion efficiency $\varepsilon_{conv}=0.937$ in \eqref{eq:2_alpha_x_total}, we obtain the profiles of $S(x)$ and $\alpha(x)$ plotted in Fig.~\ref{fig:3_tapProf_Xs_g_LP}(a). The modulation index profile is directly obtained from $\alpha(x)$ by solving the dispersion problem as discussed in Section~\ref{sec:obliqueInc_SMRS} \cite{caminita:2014,minatti:2018book}. The resulting modulated sheet reactance $X_s(x)$ is presented in Fig.~\ref{fig:3_tapProf_Xs_g_LP}(b). 
	
	\begin{figure}[t]
		\centering
		\includegraphics[width=1\columnwidth]{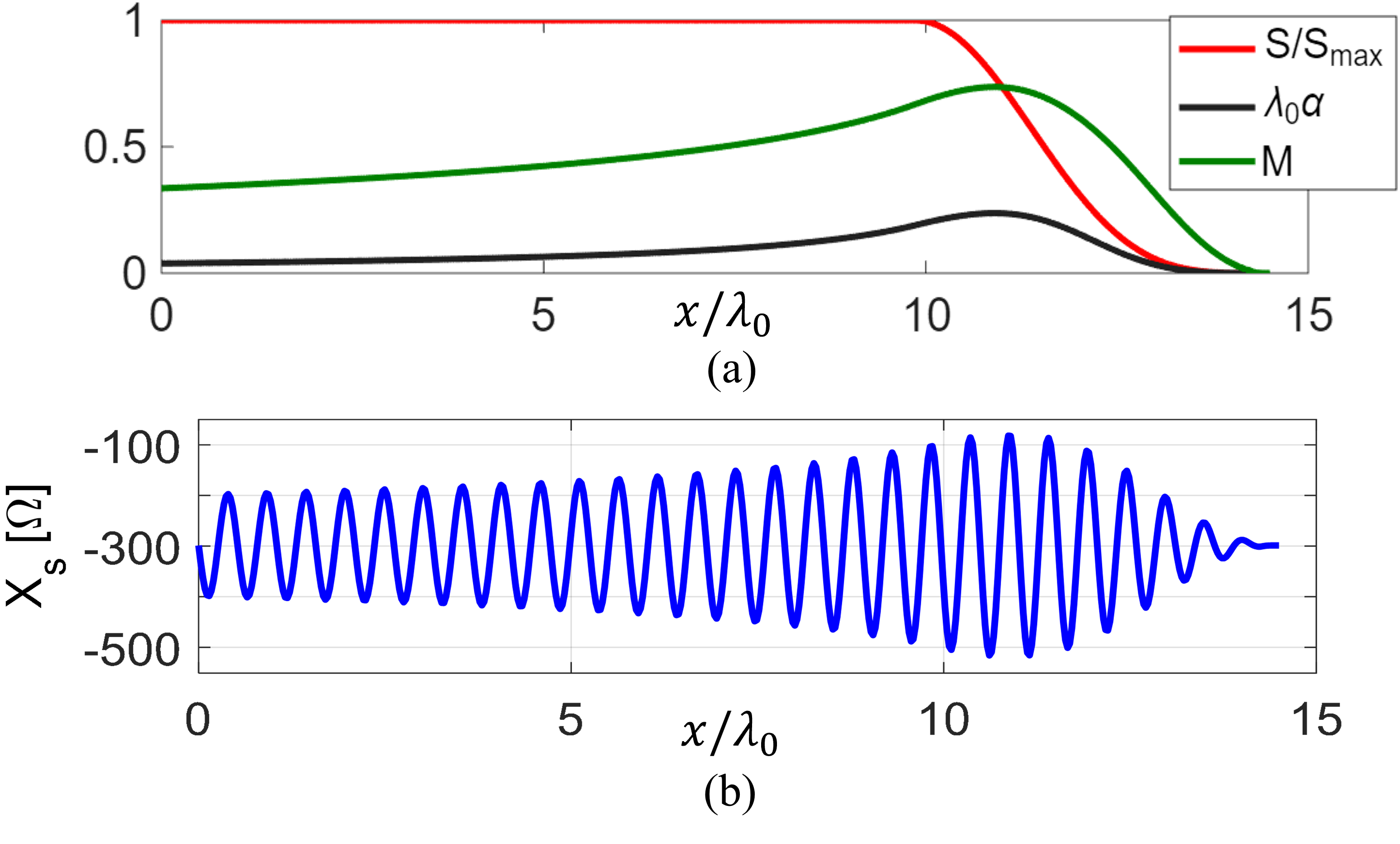}
		\caption{Spatial variation of modulated MTS along $x$. (a) Profile of imposed power density $S(x)$, required leaking factor $\alpha(x)$ and resulting spatially-varying modulation index $M(x)$. (b) Modulated sheet reactance $X_s(x)$.
		}
		\label{fig:3_tapProf_Xs_g_LP}
	\end{figure}
	
	To implement the MTS, we use the same dielectric as in the pillbox described in Section~\ref{sec:pillbox}. The geometry and dimensions of the unit cell used in the implementation are shown in Fig.~\ref{fig:3_unitCell_LP1_map}(a) \cite{patel:2011}. As observed, the cell includes two metallic strips separated by a gap that is perpendicular to the direction of modulation. The metallic strips are on a grounded dielectric substrate. Such unit cell implements a sheet impedance of the form given in \eqref{eq:2_oli_tensorX}, which produces linearly-polarized beams. The unit cell dimension is fixed to $a=1.5$ mm, which corresponds to $a\approx\lambda_0/10$ at $f_0$. This dimension respects the homogenization principle \cite{minatti:2018book}, expressed by $a \ll \lambda_0$. At the same time, this cell dimension provides a sufficiently large range of sheet impedance values while keeping higher-order modes out of the SW regime. The parameter $g$ denotes the gap spacing between metallic strips (see Fig.~\ref{fig:3_unitCell_LP1_map}(a)) and spatially varies along $x$ to produce a modulated sheet impedance. Its minimum is fixed to $g_{min}=0.1$ mm to respect manufacturing constraints. 
	
	Next, we generate a mapping function relating the parameter $g$ with an equivalent sheet impedance $Z_{s}=jX_{s}$. Here, we use a commercial full-wave eigenmode solver \cite{HFSS} to simulate the configuration depicted in Fig.~\ref{fig:3_unitCell_LP1_map}(b). We place an air box above the cell and use periodic boundary conditions (BCs) on the lateral walls. This configuration embeds the cell into a 2D periodic lattice, so that a local equivalent impenetrable impedance $Z_{op}=jX_{op}$ can be retrieved \cite{fong:2010,patel:2011}. We gradually vary $g$ and solve for the corresponding eigenfrequency $\omega$ as we stipulate different values of $\phi_x$, which denotes the phase delay across $x$ (see Fig.~\ref{fig:3_unitCell_LP1_map}(b)). For $N$ values of $g$, the simulation produces a family of $N$ dispersion diagrams relating $\phi_x$ and $\omega$. The relation between $Z_{s}$ and $g$ is obtained as follows. For a given design frequency, we recover the value of $\phi_x$ from all dispersion diagrams. This yields a function that relates $\phi_x$ and $g$. Then, we use the following expressions in the given order \cite{patel:2011},
	\begin{equation}
		k_x=\frac{\phi_x}{a}, \quad k_z=-j\sqrt{k_0^2-k_x^2}, \quad Z_{op}=\eta_0 \frac{k_z}{k_0},
		\label{eq:3_Xop_map}
	\end{equation}
	where $\eta_0$ is the free-space wave impedance. Finally, we calculate the sheet impedance $Z_s$ from $Z_{op}$ using the admittance relation $Y_s+Y_{short}^{TM}=Y_{op}$ \cite{patel:2013april}, where,
	\begin{equation}
		Y_{short}^{TM} = Y_1^{TM} \frac{1}{j\tan(k_{z_1} d)},\quad
		Y_1^{TM} = \frac{k_1}{\eta_1 k_{z_1}}.
		\label{eq:modal_transp_imped}
	\end{equation}
	Variables $\eta_1$, $k_1$ and $k_{z_1}$ correspond to the wave impedance, wavenumber and transverse fundamental wavenumber in the substrate, respectively. Since $k_x$ is continuous across the impedance sheet, it follows that $k_{z_1}=\sqrt{k_1^2-k_x^2}$. The resulting mapping function relating the sheet reactance $X_{s}$ and $g$ at the central frequency $f_0$ is plotted in Fig.~\ref{fig:3_unitCell_LP1_map}(c). One can observe that the maximum and minimum values of $X_s(x)$ plotted in Fig.~\ref{fig:3_tapProf_Xs_g_LP}(b) are within the realizable range of reactance given by $X_s(g)$ and shown in Fig.~\ref{fig:3_unitCell_LP1_map}(c). To implement the MTS, the values of $X_s(x)$ are mapped to those of $X_s(g)$. This mapping provides the spatial variation of $g$ shown in Fig.~\ref{fig:3_unitCell_LP1_map}(d).
	
	\begin{figure}[t]
		\centering
		\includegraphics[width=1\columnwidth]{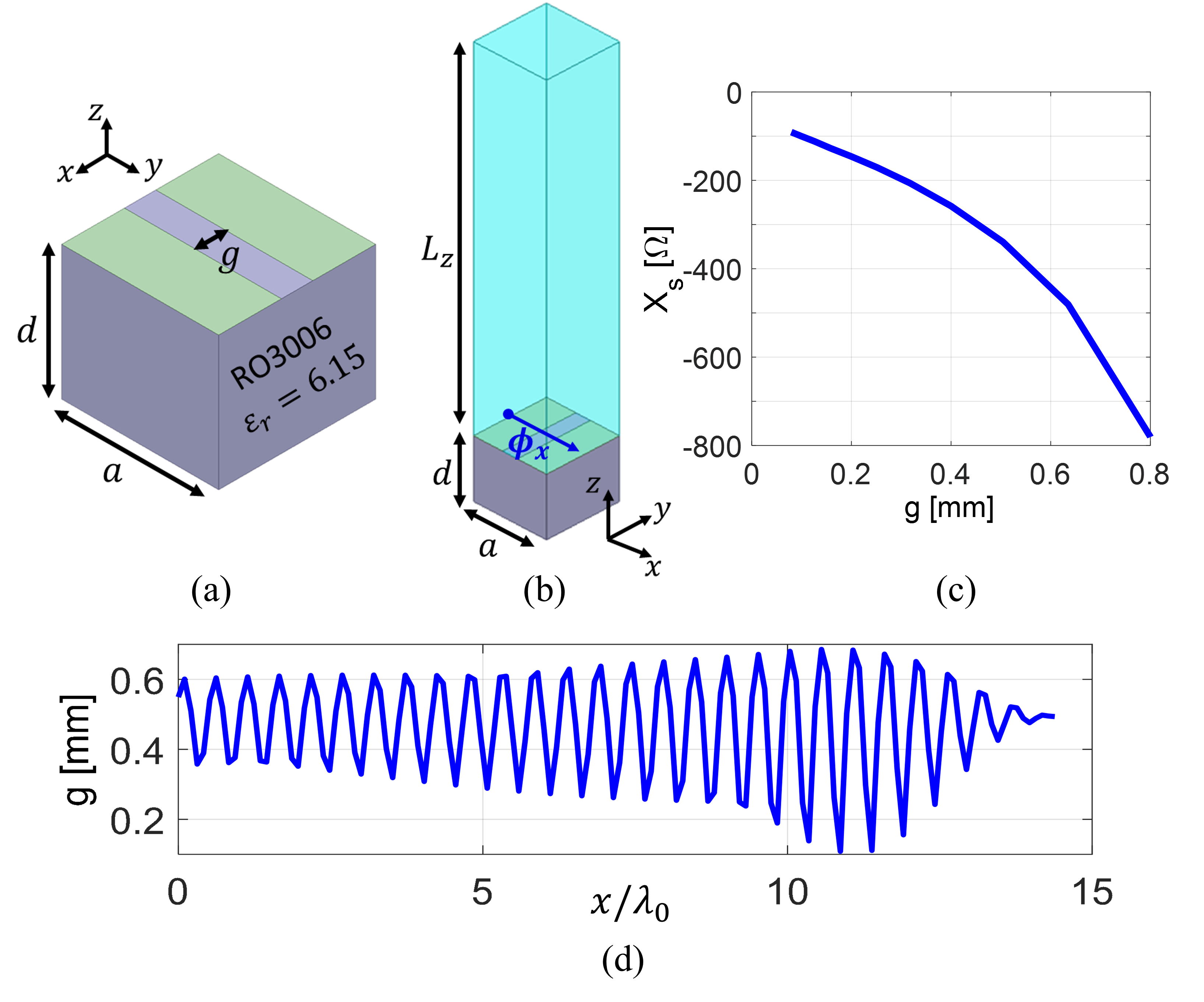}
		\caption{Unit cell geometry and its relation to an equivalent sheet reactance. (a) Geometrical parameters of the unit cell (metallic parts are in green and dielectric in gray). (b) Unit cell model used for simulation, with air box, periodic BCs on lateral walls and varying phase delay $\phi_x$ for each value of $g$. The air-box height is $L_z=10d$. (c) Mapping function relating local equivalent sheet reactance and gap spacing at $f_0=20.7$ GHz. (d) Spatial variation of gap spacing $g$.
		}
		\label{fig:3_unitCell_LP1_map}
	\end{figure}
	
	\subsection{Position of Primary Feeds}\label{sec:position_feeds}
	Now, we use the analytical model introduced in Section~\ref{sec:obliqueIncidence} to determine the position of the feeds in the pillbox bottom layer. We first use the procedure described in Section~\ref{sec:obliqueInc_quasiOptPillbox} to relate the feed position and SW propagation angle $\phi_{SW}$ on the metasurface. For the designed pillbox-fed MTS, Fig.~\ref{fig:3_phiSW_thetaPhi_LP} shows $\phi_{SW}$ as a function of the feed position. Finally, using \eqref{eq:2obli_theta_rad-1} and \eqref{eq:2obli_phi_rad-1} we can obtain the radiated beam pointing angles as functions of the feed position, as shown in Fig.~\ref{fig:3_phiSW_thetaPhi_LP}. A trade-off must be made between aperture efficiency and the achievable beam span. As the illumination becomes more oblique (larger $\phi_{SW}$) the effective active area of the MTS decreases, which in turn reduces the aperture efficiency. Conversely, a more oblique illumination increases the beam span. A large beam span can be obtained without an excessively oblique illumination by placing the offset ports at $y_{port}=\pm16$ mm (ports 2 and 3 in Fig.~\ref{fig:3_schemaPillboxMTS}(a)). With this configuration, we obtain $\phi_{SW}=11.6^\circ$ and the theoretical pointing angles $\theta_{rad}=-26^\circ$ and $\phi_{rad}=\pm51^\circ$ at $f_0$.
	
	\begin{figure}[t]
		\centering
		\includegraphics[width=1\columnwidth]{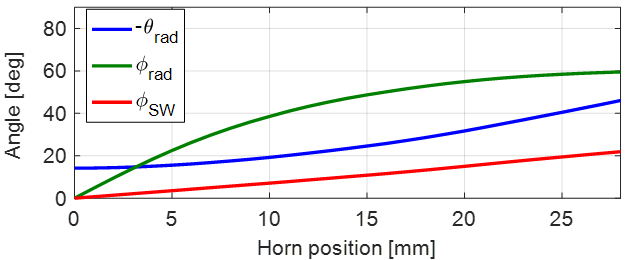}
		\caption{Calculated beam pointing angles at $f_0$ and SW propagation angle as functions of the primary feed position for the pillbox-fed MTS antenna.}
		\label{fig:3_phiSW_thetaPhi_LP}
	\end{figure}
	
	\subsection{Transition Pillbox-to-Metasurface}\label{sec:trans_pillboxMTS_LP}
	A transition is used to impedance match the TEM plane wave in the pillbox top PPW section to the TM surface wave on the MTS (see Fig.~\ref{fig:3_schemaPillboxMTS}(b)) \cite{Ruiz:2021,Ruiz:2020_eucap,Bertrand:2024}. The matching transition comprises a slot in the pillbox top PPW section and three matching unit cells in the MTS. The optimal size and position of the slot are determined using the back-to-back structure shown in Fig.~\ref{fig:3_transPPW_MTS_Sparams}(a). In the back-to-back structure, we use a non-modulated MTS with an equivalent sheet impedance $X_{s}^{av}$ (average reactance of the modulated MTS in Section~\ref{sec:MTS_design}). The gap spacing of this non-modulated MTS is denoted by $g_{av}$. For the three matching unit cells in the MTS, their gap spacings are denoted by $g_{min}$, $g_{1}$ and $g_{2}$. A linear progression from $g_{min}$ to $g_{av}$ is followed to determine $g_{1}$ and $g_{2}$. We use periodic BCs on the lateral walls of the back-to-back structure and radiation BCs on the remaining faces (bottom face is a Perfect Electric Conductor, PEC). Finally, a full-wave solver \cite{CST} is used to find the optimal slot size and position that maximize the transmission coefficient $|S_{21}|$ of the back-to-back structure. The prefixed and optimized dimensions of the structure are given in Table~\ref{table:transition_pillboxMTS}, while the resulting magnitude of the S-parameters is shown in Fig.~\ref{fig:3_transPPW_MTS_Sparams}(b). 
	
	\begin{figure}[t]
		\centering
		\includegraphics[width=1\columnwidth]{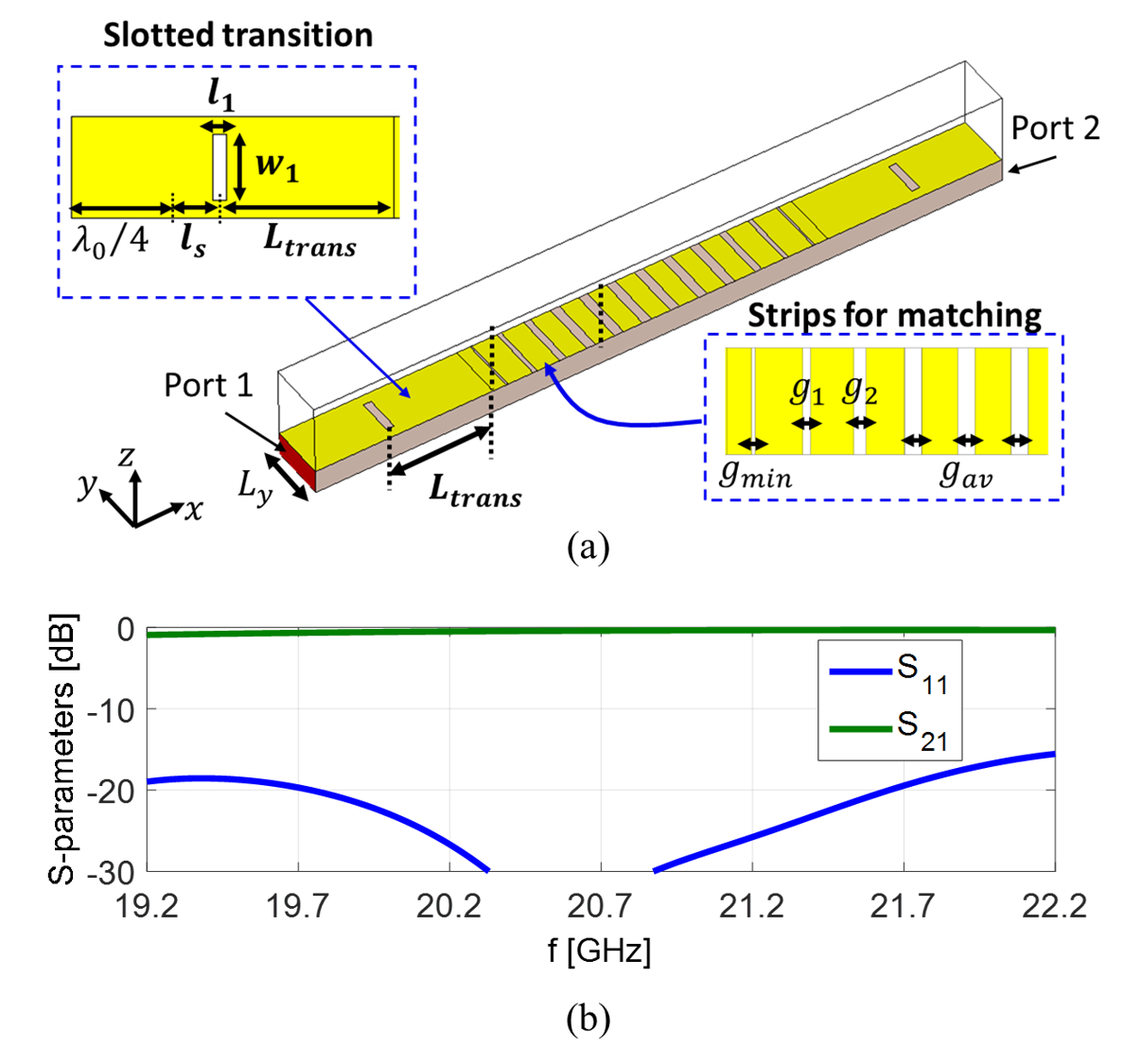}
		\caption{Design of transition to impedance match pillbox top PPW section and modulated MTS. (a) Back-to-back structure used to optimize size and position of the slot in the PPW section (optimized parameters are shown in bold). (b) Magnitude of S-parameters for the optimized matching transition.
		}
		\label{fig:3_transPPW_MTS_Sparams}
	\end{figure}
	
	{\setlength{\tabcolsep}{4pt}
		\begin{table}[t]
			\centering
			\caption{Parameters in millimeters of the matching transition pillbox-to-MTS shown in Fig.~\ref{fig:3_transPPW_MTS_Sparams}(a).}
			\begin{tabular}{ |c|c|c|c|c|c|c|c|c|c|c| }
				\hline
				$g_{av}$ & $g_{min}$ & $g_{1}$ & $g_{2}$ & $L_y$ & $L_{trans}$ & $l_s$ & $l_1$ & $w_1$ \\
				\hline
				$0.48$ & $0.1$ & $0.23$ & $0.35$ & $3$ & $5.83$ & $0.71$ & $0.41$ & $1.93$ \\
				\hline
			\end{tabular}
			\label{table:transition_pillboxMTS}
		\end{table}
	}
	
	\begin{figure}[t]
		\centering
		\includegraphics[width=1\columnwidth]{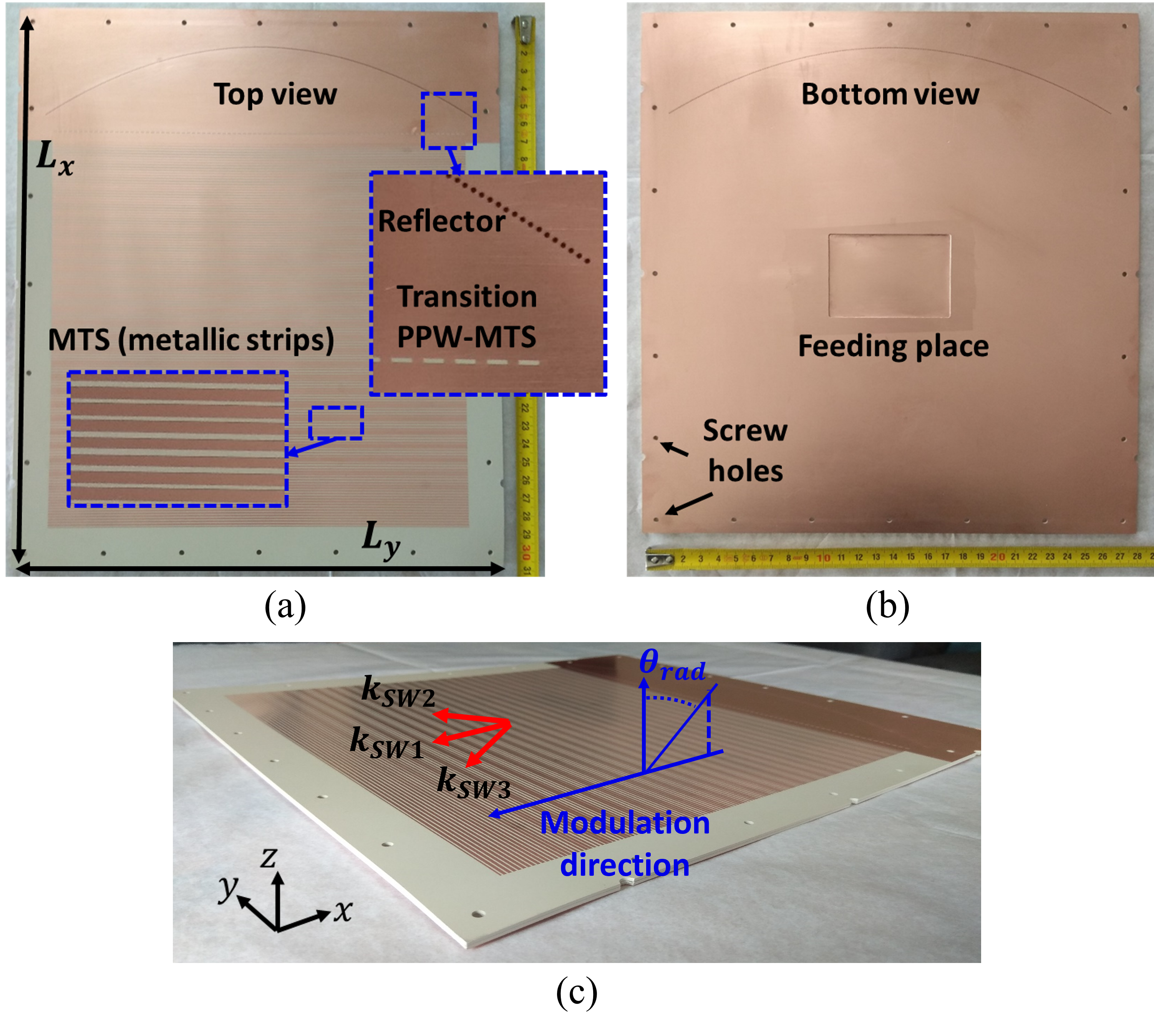}
		\caption{Prototype of pillbox-fed multibeam modulated MTS antenna. (a) Top and (b) bottom view of the antenna. Screw holes are included in the antenna frame for assembly with the feeding network (see Appendix~\ref{sec:transitionWg}). (c) Perspective view of the antenna including the coordinate system used for measurements and the SW wavevectors for each port. 
		}
		\label{fig:fabricated_MTSantenna}
	\end{figure}
	
	\section{Antenna Prototype and Experimental Results}\label{sec:pillboxMTS_LP_simu}
	The fabricated pillbox-fed multibeam MTS antenna is shown in Fig.~\ref{fig:fabricated_MTSantenna}, having dimensions $L_x=31.2$ cm and $L_y=28.4$ cm. The manufacturing process is described in Appendix~\ref{sec:pillboxMTS_LP_manuf_process}. The antenna is fed from the bottom layer by an all-metal feeding network that excites the pillbox primary feeds \cite{Bertrand:2024}. Details about the design and fabrication of the feeding network and its assembly with the MTS antenna are given in Appendix~\ref{sec:transitionWg}. Before discussing the experimental results, let us clarify the coordinate system used to describe the beam direction. As explained in Section~\ref{sec:pillboxMTS_LPantenna_design}, the MTS is designed to radiate multiple beams with elevation angles $\theta_{rad}<0$. This definition of $\theta_{rad}$ considers the coordinate system shown in Fig.~\ref{fig:2_obli_conceptFig}, where the illuminating SW wavevectors have a positive $x$-component. For the experimental results, however, we use the coordinate system shown in Fig.~\ref{fig:fabricated_MTSantenna}(c). In this case, the various illuminating plane SWs have a negative $x$-component and the radiated beams exhibit elevation angles $\theta_{rad}>0$.
	
	\begin{figure}[t]
		\centering
		\includegraphics[width=1.0\columnwidth]{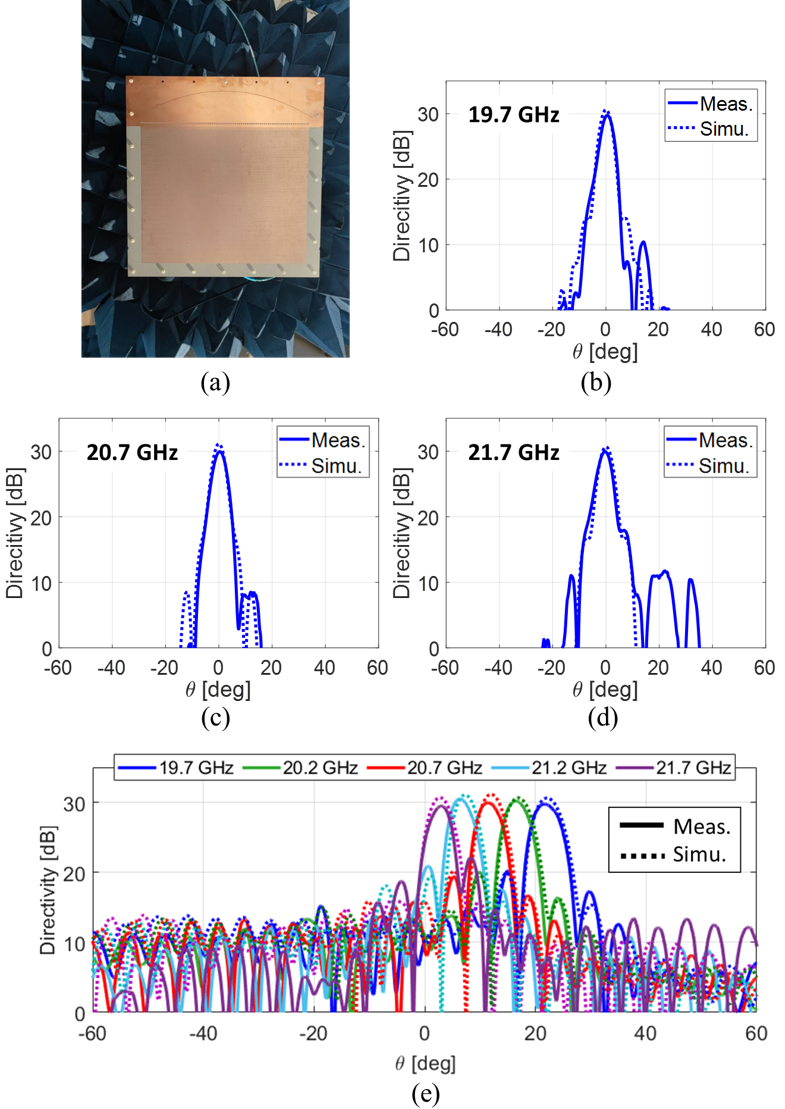}
		\caption{Experimental set-up and results for the fabricated MTS antenna. (a) Set-up to measure radiation performance. Measured and simulated directivity patterns when port 1 is active for H-plane at (b) 19.7 GHz, (c) 20.7 GHz and (d) 21.7 GHz. (e) Measured and simulated directivity patterns when port 1 is active for E-plane for several frequencies over the 19.7-21.7 GHz band.}
		\label{fig:3_Epatt_Hpatt_LP1_simuVsMeas}
	\end{figure}
	
	The antenna has been measured at the Instituto Nacional de T\'ecnica Aeroespacial (INTA), Madrid, Spain. The radiation measurement set-up is shown in Fig.~\ref{fig:3_Epatt_Hpatt_LP1_simuVsMeas}(a). Directivity patterns in the H-plane for port 1 (central port) are shown in Fig.~\ref{fig:3_Epatt_Hpatt_LP1_simuVsMeas}(b)-(d) for several frequencies over the 19.7-21.7 GHz band. Fig.~\ref{fig:3_Epatt_Hpatt_LP1_simuVsMeas}(e) shows the directivity patterns in the E-plane for several frequencies over the band of interest when port 1 is active. The agreement between the simulated and measured results is good in both E and H-plane. In addition, it can be observed that the beam symmetry is preserved over the band of interest. We note that the measured and simulated cross-polarization discrimination is below $-30$ dB for all the ports over the entire band.
	
	\begin{figure}[t]
		\centering
		\includegraphics[width=1\columnwidth]{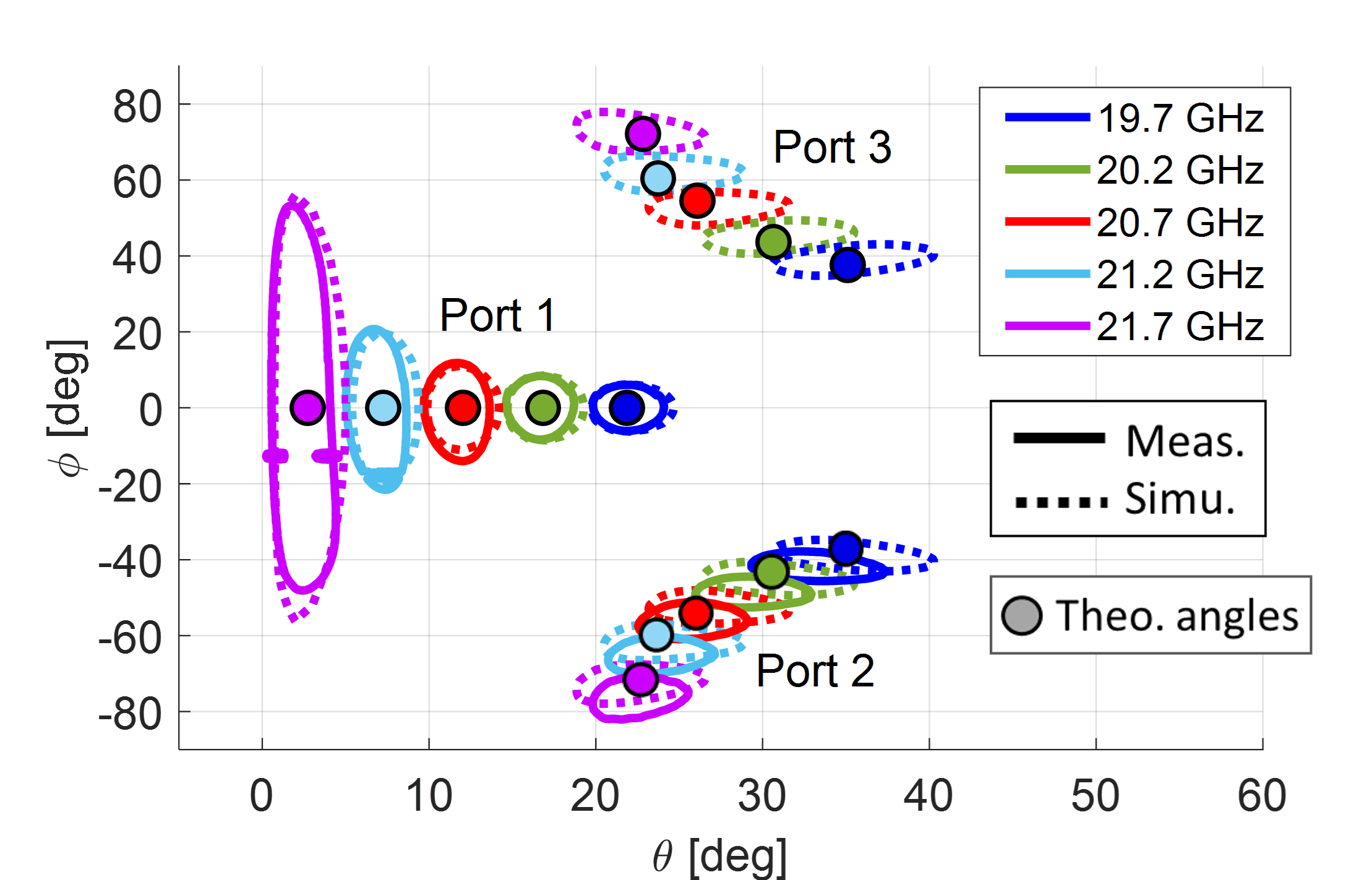}
		\caption{Measured and simulated -3-dB contour diagram for several ports and frequencies over the 19.7-21.7 GHz band. Simulated contours for port 3 are included to illustrate the symmetry with port 2. Solid points represent the theoretical beam pointing angles calculated using expressions \eqref{eq:2obli_theta_rad-1} and \eqref{eq:2obli_phi_rad-1} for the various frequencies.
		}
		\label{fig:3_3dB_diag_LP1_simuVsMeas}
	\end{figure}
	
	\begin{figure}[t]
		\centering
		\includegraphics[width=1\columnwidth]{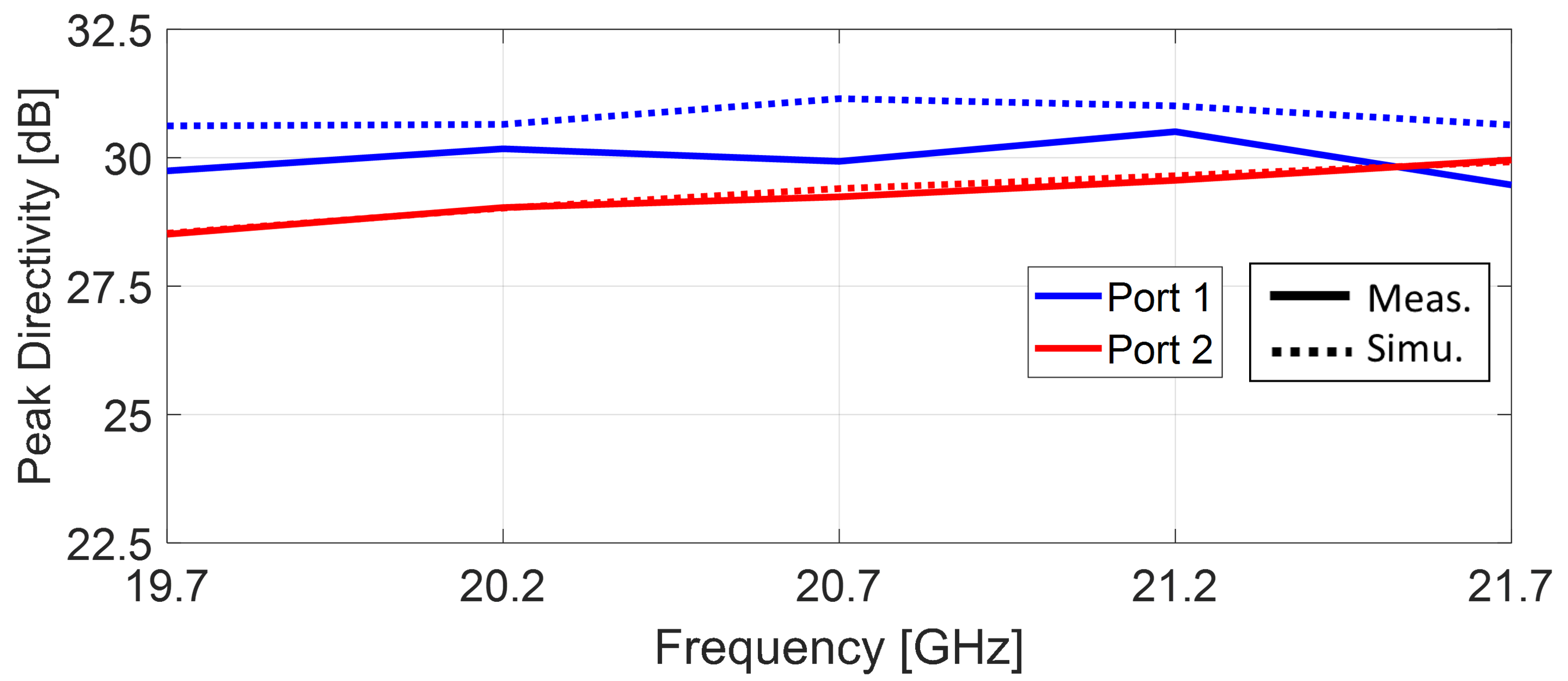}
		\caption{Measured and simulated peak directivity over the 19.7-21.7 GHz band for port 1 and 2 (port 3 provides practically the same result as port 2).}
		\label{fig:3_Dpeak_freq_LP1_simuVsMeas}
	\end{figure}
	
	The beam switching and frequency scanning performance of the antenna is illustrated in Fig.~\ref{fig:3_3dB_diag_LP1_simuVsMeas}. Measured and simulated \mbox{-3-dB} contours are shown for all ports and several frequencies over the 19.7-21.7 GHz band (measurements for port 3 have been omitted due to symmetry with port 2). For reference, we include in Fig.~\ref{fig:3_3dB_diag_LP1_simuVsMeas} the theoretical beam pointing angles calculated using the closed-form expressions derived in Section~\ref{sec:obliqueIncidence}. The angular difference between measured and theoretical results remains below $1.6^\circ$ in $\theta$ and $1.5^\circ$ in $\phi$ for all ports and over the band of interest. When switching ports at $f_0$, the measured beam span is $\theta_{rad}\in[11.5^\circ,25.5^\circ]$ and $\phi_{rad}\in[-52^\circ,+52^\circ]$ (see red contours in Fig.~\ref{fig:3_3dB_diag_LP1_simuVsMeas}). With frequency scanning, the measured beam span becomes $\theta_{rad}\in[3^\circ,35^\circ]$ and $\phi_{rad}\in[-76^\circ,+76^\circ]$. As observed in Fig.~\ref{fig:3_3dB_diag_LP1_simuVsMeas}, there is a good agreement between measured, simulated and theoretical results when using both frequency scanning and beam switching. Fig.~\ref{fig:3_Dpeak_freq_LP1_simuVsMeas} shows the measured and simulated peak directivity for ports 1 and 2 (port 3 provides practically the same results as port 2) over the 19.7-21.7 GHz band. A maximum difference of $1.3$ dB is observed between measured and simulated peak directivity for port 1. This difference is attributed to a fabrication error in the central primary feed, which was dented during the carving of the feeding network receptacle shown Fig.~\ref{fig:fabricated_MTSantenna}(b). The agreement between measured and simulated peak directivity for port 2 is very good over the band of interest. The antenna provides a measured maximum directivity of 30.5 dB for port 1 and 30 dB for port 2 and 3. 
	
	\begin{figure}[t]
		\centering
		\includegraphics[width=1\columnwidth]{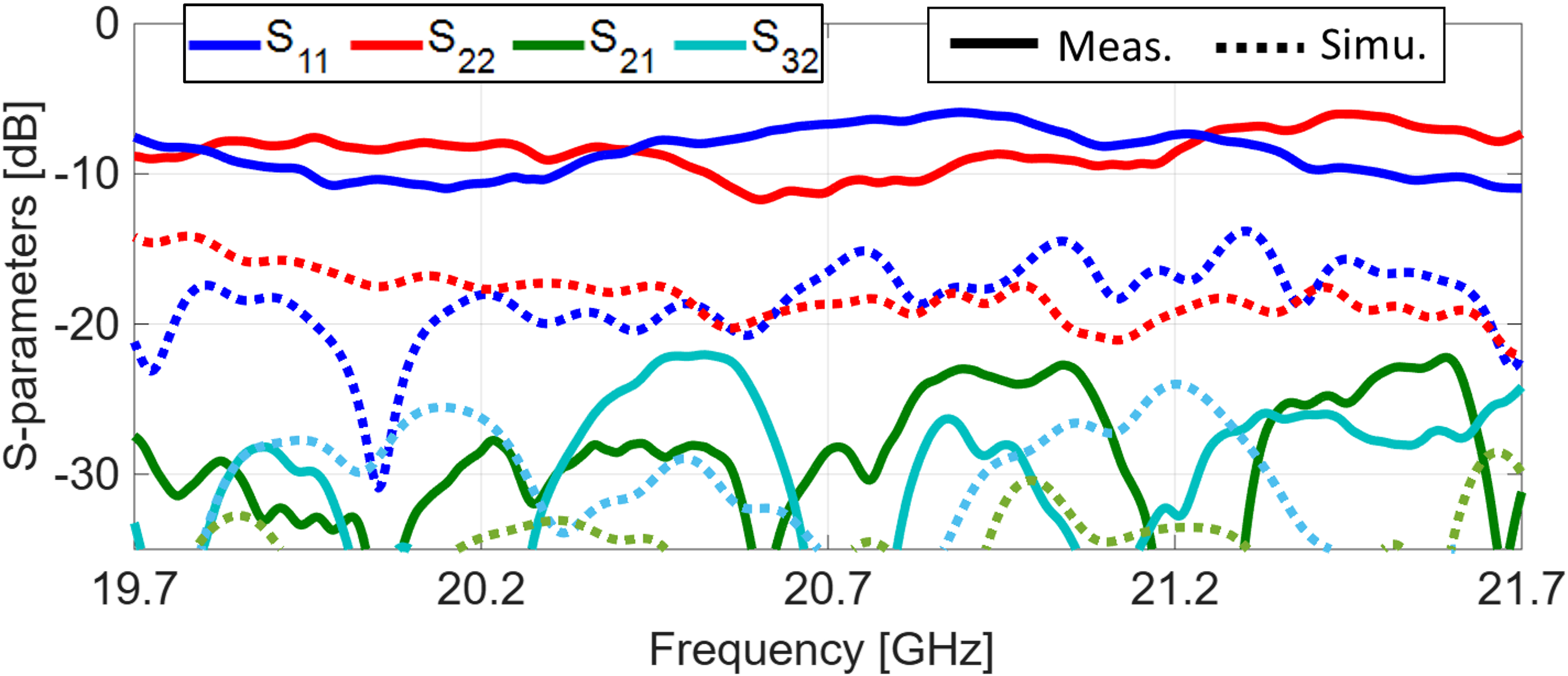}
		\caption{Measured and simulated magnitudes of the S-parameters for the fabricated multibeam MTS antenna.}
		\label{fig:3_Sparams_LP1_measuredSimu}
	\end{figure}
	
	The measured maximum realized gain is 25.3 dBi for port 1 and 24.7 dBi for ports 2 and 3, which is 5-6 dB below the measured directivity. While the simulated realized gain for all ports is 3-4 dB below the measured directivity over the 19.7-21.7 GHz band, the measured realized gain is lower. The 3-4 dB difference observed in simulation is expected due to propagation losses in the lower and upper substrates of the antenna. However, the reduction in measured realized gain is attributed to a misalignment between the feeding network's output waveguides (see Appendix~\ref{sec:transitionWg}) and the pillbox primary feeds \cite{Bertrand:2024}. This issue is quantified in Fig.~\ref{fig:3_Sparams_LP1_measuredSimu}, which shows the measured and simulated magnitudes of the S-parameters (due to symmetry with port 3, only the results for port 2 are included). Although the coupling between ports is below $-20$ dB over the band, both measured $S_{11}$ and $S_{22}$ magnitudes show a poor impedance match. More details about this matching problem are given in Appendix~\ref{sec:transitionWg}. Beyond the matching issue, we note that the main purpose of the fabricated antenna is to experimentally validate the beam pointing angle prediction.

	\section{Conclusion}\label{sec:conclusion}
	Exciting modulated metasurfaces (MTSs) by plane waves with different directions is a versatile method to design multibeam antennas with wide scanning capabilities. Both plane-wave launchers and MTSs can be engineered to achieve compact, highly directive antennas. In this paper, we focus on MTSs that are sinusoidally modulated along one direction in a Cartesian coordinate system. We show that this class of MTSs can be effectively exploited to realize multibeam antennas with different geometries and form factors. 
	
	First, we have solved the problem of plane surface waves (SWs) that propagate obliquely to an MTS modulation direction. The derived closed-form expressions provide the beam pointing angles in azimuth and elevation as functions of the SW propagation direction. We have also addressed the case of discontinuous media. This analysis is relevant in most practical situations, where the SW is generated by an external plane-wave launcher. We have included several design examples that combine unidirectionally sinusoidally modulated MTSs with different plane-wave launchers. Rectangular and circular antenna form factors have been investigated. Numerical results for these design examples corroborate the generality and accuracy of the presented formulation.
	
	Then, the derived formulation has been applied to design a quasi-optically-fed multibeam MTS antenna. A pillbox beamformer has been used as plane-wave launcher. Pillbox beamformers are double-layer structures that integrate multiple primary feeds and generate plane waves with different directions. When combined with modulated MTSs, this multi-directional plane-wave illumination produces multiple directive beams. The pillbox-fed multibeam antenna has been fabricated and experimentally characterized. The measured results validate the proposed formulation for designing low-profile, highly directive multibeam MTS antennas with beam scanning and beam switching. The prototype exhibits a beam span of [3\textdegree,35\textdegree] in elevation and [-76\textdegree,+76\textdegree] in azimuth over the 19.7-21.7 GHz band. In addition, the antenna provides 30.5 dB of maximum directivity and a 17.5\% fractional -3 dB directivity bandwidth. Due to an unanticipated misalignment between the fabricated antenna and its feeding network, the maximum realized gain reduces to 25.3 dBi. Potential applications of the proposed multibeam concept include 5G/6G networks as well as communications and sensing for satellite and military systems.

	
	\appendices
	
	\section{Manufacturing Process of Multibeam Metasurface Antenna}\label{sec:pillboxMTS_LP_manuf_process}
	Fig.~\ref{fig:3_pillboxMTS_sub1_pins_transHole} shows the CAD model of the fabricated multibeam MTS antenna. The primary feeds and pillbox parabolic reflector are implemented using substrate integrate waveguide (SIW) technology \cite{Deslandes:2006}. The diameter and periodicity of the metallized vias are $d_{p}=0.7$ mm and $p_{p}=0.9$ mm, respectively (see Fig.~\ref{fig:3_pillboxMTS_sub1_pins_transHole}(b)). As explained in Section~\ref{sec:pillbox}, the pillbox focal distance is intentionally short to minimize the propagation losses in the dielectric, while ensuring a symmetric pencil beam. In order to accommodate the feeding network, which is described in Appendix~\ref{sec:transitionWg}, a cavity is carved in the bottom layer as shown in Fig.~\ref{fig:3_pillboxMTS_sub1_pins_transHole}(c).
	
	\begin{figure}[t]
		\centering
		\includegraphics[width=1\columnwidth]{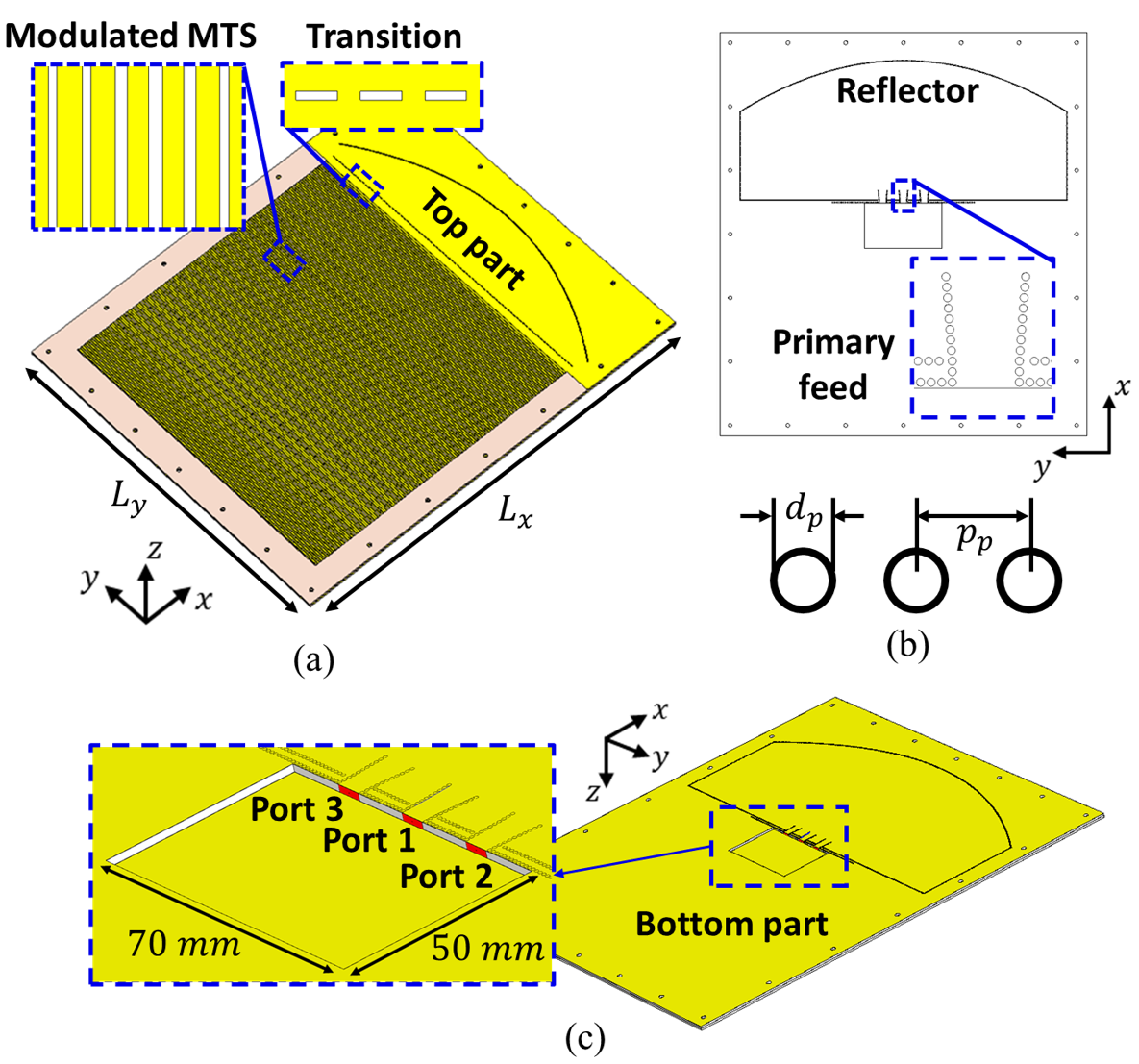}
		\caption{CAD model of fabricated multibeam MTS antenna. (a) Top part of the antenna with details of modulated MTS and pillbox-to-MTS transition. (b) Bottom view of the antenna with details of the SIW primary feeds and parabolic reflector. (c) Bottom part of the antenna with details of the cavity to accommodate the feeding network.
		}
		\label{fig:3_pillboxMTS_sub1_pins_transHole}
	\end{figure}

	\begin{figure}[t]
		\centering
		\includegraphics[width=1\columnwidth]{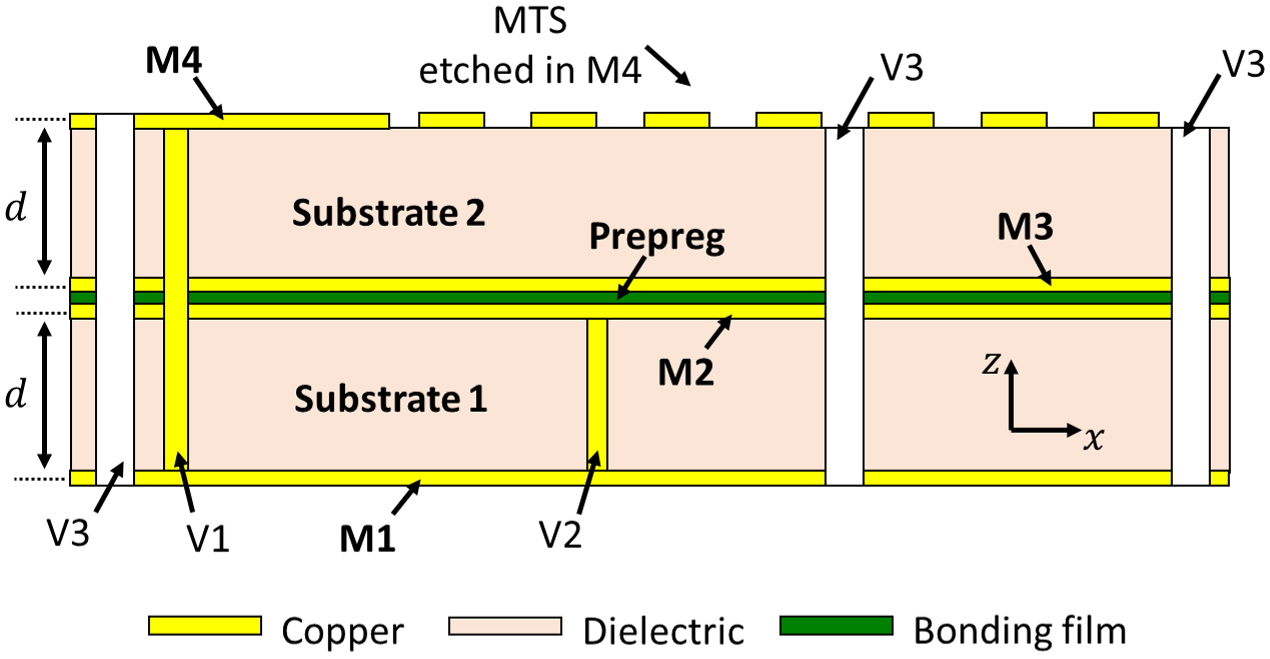}
		\caption{Multilayer PCB stack-up and vias used for the fabrication of the multibeam MTS antenna.
		}
		\label{fig:3_fabric_LP_pictures}
	\end{figure}
	
	The antenna was fabricated using standard multilayer PCB technology at the company A \& P Lithos \cite{lithos:web}. Fig.~\ref{fig:3_fabric_LP_pictures} shows a section view of the antenna with details of metallic, dielectric and prepreg layers. The metallic and fixation vias through the layers are also displayed. As mentioned in Section~\ref{sec:pillboxMTS_LPantenna_design}, the substrate is Rogers RO3006 with a copper thickness $d_{metal}=0.17$ $\mu$m. The prepreg material used to glue the two substrate layers is CuClad 6250 Bonding Film with $d=38$ $\mu$m, $\varepsilon_r=2.32$ and $\tan \delta=0.0015$. Since the bonding film is extremely thin with respect to the substrate, its presence has a negligible impact on the power coupling between bottom and top layer. In addition, the thinness of the bonding film prevents higher-order PPW modes from exciting in the gap between layers (green region in Fig.~\ref{fig:3_fabric_LP_pictures}). 
	
	As for the manufacturing procedure, the vias in the lower substrate (V2) are drilled first, connecting the copper layers M1 and M2. These are the vias illustrated in Fig.~\ref{fig:3_pillboxMTS_sub1_pins_transHole}(b). Next, the cavity to accommodate the feeding network is carved in the bottom layer (see Fig.~\ref{fig:3_pillboxMTS_sub1_pins_transHole}(c)) and the coupling slots (see Fig.~\ref{fig:3_schemaPillboxMTS}(a)) are etched in M2. Before joining the two layers, the coupling slots are also etched in M3 at the top layer. After aligning and gluing both layers, the remaining vias are drilled. Specifically, remaining vias are V1, which join all metallic layers and form the parabolic reflector, and the screw holes V3, which affix the antenna to the support and feeding network. Finally, the pillbox-to-MTS transition and modulated MTS motifs are etched on the topmost metallic layer M4. The resulting antenna prototype is shown in Fig.~\ref{fig:fabricated_MTSantenna}.

	\section{Design of Feeding Network}\label{sec:transitionWg}
	The feeding network of the fabricated antenna is formed from all-metal stepped rectangular waveguides \cite{cano:2015}, as shown in Fig.~\ref{fig:3_pillboxMTS_transition_yCut}. The feeding network matches the input impedance of the antenna's primary feeds to that of standard WR-42 waveguides. For practical reasons, the feeding network is integrated into the antenna support \cite{Bertrand:2024}. Alternative feeding solutions for double-layer antenna designs similar to ours use coaxial SMA \cite{Yurduseven:2020} or mini-SMP connectors \cite{Tekkouk:2015cross}. However, the proposed all-metal feeding approach presents two main advantages. First, connector-based alternatives require a matching transition generally consisting of co-planar waveguides (CPWs). Using all-metal waveguides avoids substrate-based complex transitions, reducing losses. Second, when comparing to models as the one in \cite{Yurduseven:2020}, our feeding network is placed at the bottom part of the antenna. Therefore, it minimizes the feeding interaction with the radiated fields from the antenna aperture.
	
	\begin{figure}[t]
		\centering
		\includegraphics[width=1\columnwidth]{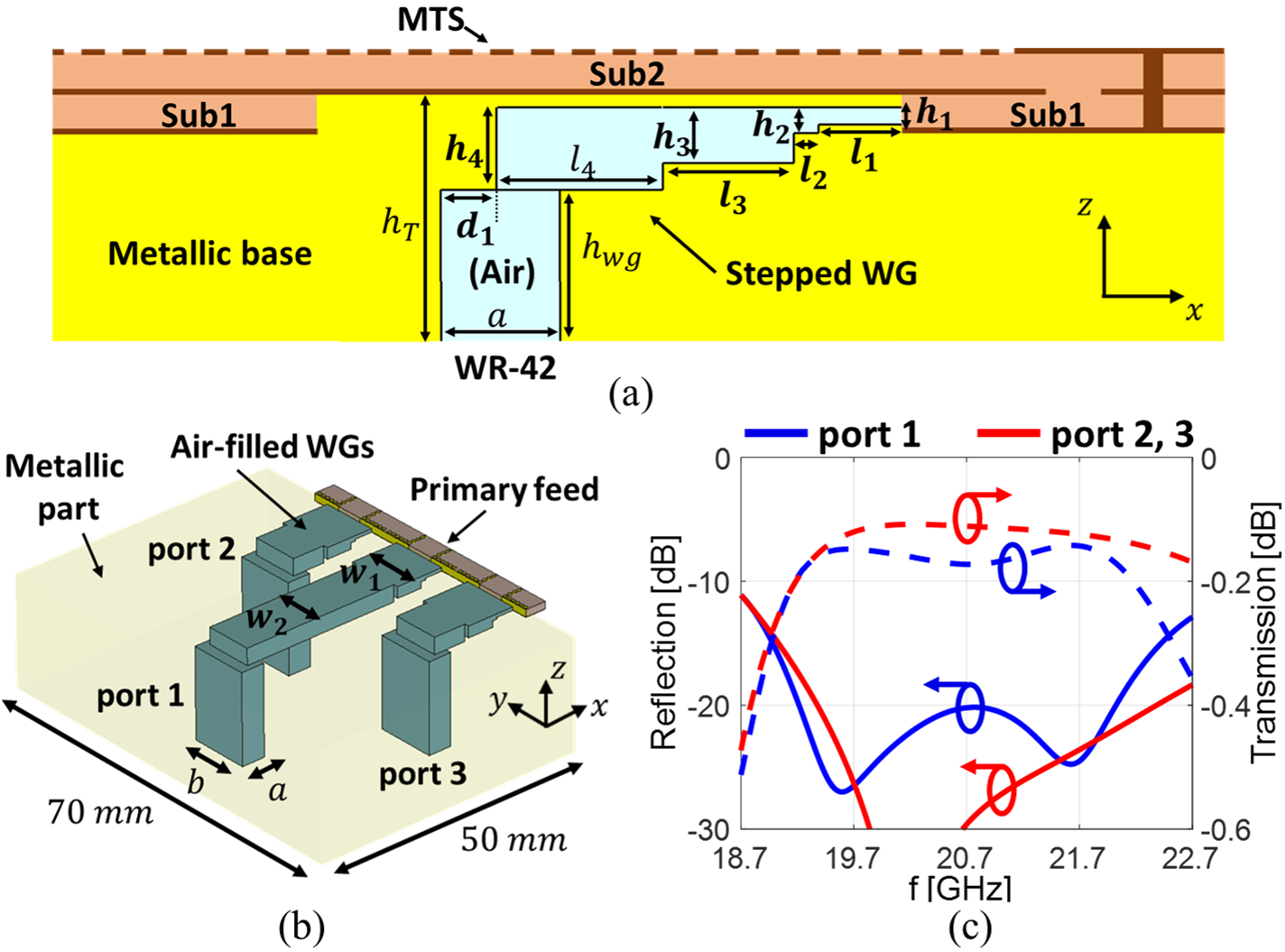}
		\caption{Feeding network to excite the fabricated multibeam MTS antenna. (a) Section view of overall antenna assembly including double-layer pillbox-fed MTS antenna, one stepped waveguide of feeding network and antenna metallic base. $Sub1$ and $Sub2$ correspond to the substrate in the pillbox lower and upper layer, respectively. (b) Stepped waveguides for all ports in the all-metal feeding network. In (a) and (b), optimized parameters are shown in bold. (c) Simulated reflection and transmission coefficients for all ports of all-metal feeding network.
		}
		\label{fig:3_pillboxMTS_transition_yCut}
	\end{figure}
	
	\begin{table}[t]
		\centering
		\caption{Fixed parameters in millimeters of all-metal feeding network (see Fig.~\ref{fig:3_pillboxMTS_transition_yCut}(a)-(b)). }
		\begin{tabular}{ |c|c|c|c|c|c|c| }
			\hline
			$a$ & $b$ & $h_T$ & $h_{wg}$ & $l_4$ (port 1) & $l_4$ (port 2-3) \\
			\hline
			$4.318$ & $10.668$ & $19.8$ & $16.127$ & $29$ & $6$ \\
			\hline
		\end{tabular}
		\label{table:3_steppedWG_values_fixed}
	\end{table}
	
	\begin{table}[t]
		\centering
		\caption{Optimized parameters in millimeters of all-metal feeding network (bold parameters in Fig.~\ref{fig:3_pillboxMTS_transition_yCut}(a)-(b)).}
		\begin{tabular}{ |c|c|c|c|c| }
			\hline
			$h_1$ & $h_2$ & $h_3$ & $h_4$ & $l_1$ \\
			\hline
			$0.61$ & $0.921$ & $2.011$ & $2.973$ & $3.201$ \\
			\hline
			$l_2$ & $l_3$ & $w_1$ & $w_2$ & $d_1$ \\
			\hline
			$0.896$ & $4.752$ & $10.206$ & $8.543$ & $2.033$ \\
			\hline
		\end{tabular}
		\label{table:3_steppedWG_values_opt}
	\end{table}
	\begin{figure}[!t]
		\centering
		\includegraphics[width=1\columnwidth]{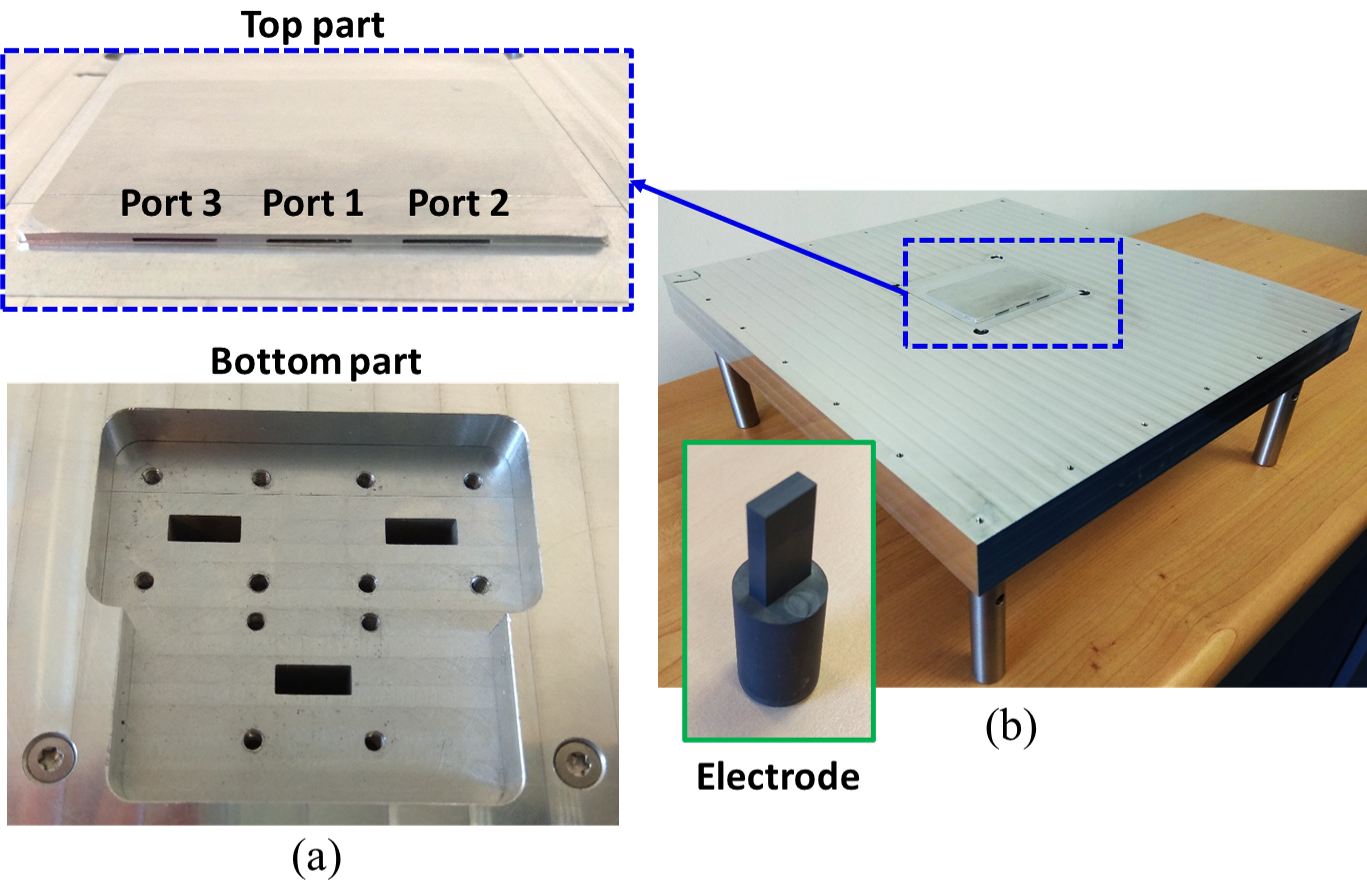}
		\caption{Fabricated all-metal feeding network and antenna base. (a) Top and bottom parts of feeding network. (b) Antenna base with embedded feeding network (the inset shows one of the electrodes used to carve the feeding network's stepped waveguides).
		}
		\label{fig:3_steppedWG_fabricated}
	\end{figure}
	
	The geometry and dimensions of the feeding network are detailed in Fig.~\ref{fig:3_pillboxMTS_transition_yCut}(a)-(b). Fig.~\ref{fig:3_pillboxMTS_transition_yCut}(a) shows a section view of the antenna integrating the feeding network. In particular, we show the all-metal stepped waveguide that matches the offset ports (port 2 and 3). The dimensions of the stepped waveguides are obtained as follows. We first fix the parameters given in Table~\ref{table:3_steppedWG_values_fixed}. Then, the remaining parameters are optimized using a commercial full-wave solver \cite{CST} to maximize the power transfer from WR-42 waveguides to the antenna's primary feeds. The optimized parameters are listed in Table~\ref{table:3_steppedWG_values_opt}. The simulated reflection and transmission coefficients for the all-metal feeding network over the 19.7-21.7 GHz band are shown in Fig.~\ref{fig:3_pillboxMTS_transition_yCut}(c). To ensure sufficient separation to accommodate the WR-42 flanges, the central waveguide (corresponding to port 1) is extended, as shown in Fig.~\ref{fig:3_pillboxMTS_transition_yCut}(b). The stepped waveguides for port 2 and 3 are identical. The fabricated feeding network is shown in Fig.~\ref{fig:3_steppedWG_fabricated}(a). The stepped waveguides have been realized in an aluminum block using electrical discharge machining (EDM). We have used as many electrodes (see inset in Fig.~\ref{fig:3_steppedWG_fabricated}(b)) as there are steps in the waveguides. Fig.~\ref{fig:3_steppedWG_fabricated}(a) shows the top and bottom parts of the fabricated feeding network. The feeding network is ultimately embedded into the antenna metallic base/support, as shown in Fig.~\ref{fig:3_steppedWG_fabricated}(b).
	
	The impedance‑matching issue mentioned in the main text arises from a misalignment between the output ports of the feeding network and the pillbox primary feeds (see Fig.~\ref{fig:3_pillboxMTS_transition_yCut}(b)). The fabricated MTS antenna is very thin (its total thickness is below 3 mm) and is not perfectly flat. This slight non‑flatness prevents the antenna receptacle in the bottom layer (see Fig.~\ref{fig:3_pillboxMTS_sub1_pins_transHole}(c)) from fully settling onto the top part of the feeding network embedded in the base (see Fig.~\ref{fig:3_steppedWG_fabricated}(a), top part). This causes the mentioned misalignment. As listed in Table~\ref{table:3_steppedWG_values_opt}, the height of the last stepped‑waveguide section in the feeding network ($h_1$) is below 1 mm. Therefore, even a small vertical offset, such as a few hundreds of $\mu m$, between the final waveguide section and a pillbox primary feed leads to a relatively large misalignment.

	\section*{Acknowledgment}
	The authors would like to thank Prof.~Stefano Maci (University of Siena, Italy) for his insightful comments on the beam-direction analysis, and Jos\'e Roberto Rodr\'iguez Amor (Instituto Nacional de T\'ecnica Aeroespacial, INTA, Madrid, Spain) for his thorough work in measuring the antenna. They would also like to thank Laurent Cronier and Christophe Guitton (IETR, France) for fabricating the all‑metal feeding network and the antenna’s metallic support.

	\bibliographystyle{IEEEtran}
	\bstctlcite{IEEEexample:BSTcontrol}
	\bibliography{IEEEabrv,biblio_2Dpredict_LPpillboxMTS}

\end{document}